\documentclass[preprint,aps,nofootinbib]{revtex4}
\usepackage{float}
\usepackage{graphicx}
\input{epsf.sty}
\usepackage{epsfig}
                                                                             
\def\lsim{\mathrel{\raise.3ex\hbox{$<$\kern-.75em\lower1ex\hbox{$\sim$}}}}
\def\gsim{\mathrel{\raise.3ex\hbox{$>$\kern-.75em\lower1ex\hbox{$\sim$}}}}
                                                                             
\def\singleandabitspaced{\baselineskip=\normalbaselineskip\multiply
    \baselineskip by 150\divide\baselineskip by 100}

\begin{document}
                                                                             
                                                                             
\vskip 0.4cm
                                                                             
\title{Effective Quark Antiquark Potential in the Quark Gluon Plasma from Gravity Dual Models}
\author{Oleg Antipin \footnote{oaanti02@iastate.edu}, Piyabut Burikham \footnote{piyabut@iastate.edu}, Jun Li \footnote{junli@iastate.edu}}
\vspace*{0.5cm}
\affiliation{Department of Physics and Astronomy, Iowa State University, \\
        Ames, Iowa 50011, USA \\} 
                                                                             
\date{\today}
                                                                             
\vspace*{2.0cm}

\begin{abstract}
We estimate the effective heavy quark and antiquark potential in the quark gluon plasma using the gravity dual theory.  Two models are considered: $AdS_{5}$ and Sakai-Sugimoto model.  The effective potential, obtained by using the rotating fundamental open-string configurations, has the angular momentum dependence which generalizes the static central potential.  For zero angular momentum case, we obtain asymptotic form of the potential for general metric $n$, and for both gravity dual models where $n=3,4$.  The mass dependence of the potential is derived at the leading order together with its temperature dependence.  Motivated by the asymptotic form for zero angular momentum state, we fit the effective potential for $J=1,2$ states in the binding region.  The fitting parameters are found to be functions of temperature.  Finally, we discuss the differences and similarities of the effective potential between the two gravity dual models.  An interesting result is that position of the minimum of the potential is determined only by angular momentum and independent of the temperature.  

\end{abstract}

\maketitle

                                                                             
\newpage

\setcounter{page}{2}
\renewcommand{\thefootnote}{\arabic{footnote}}
\setcounter{footnote}{0}
\singleandabitspaced
                                                                             
\section{Introduction}

$J/\psi$, a $c\bar{c}$ bound state, is a good probe for the formation as well as the properties of the quark gluon plasma~(QGP).  At low temperature, the interaction between $c$ and $\bar{c}$ can be phenomenologically described as Coulomb-like potential $1/r$.  Because of confinement, one would expect there is additional linear potential $\sigma r$ which is dominant at long distance.  When the temperature increases, the string tension $\sigma$ decreases.  At the critical temperature where deconfinement phase transition takes place, one would expect that $\sigma\simeq 0$.  In QGP, the existence of deconfined quarks and gluons would modify the interaction between quark and antiquark such that the static potential 
\begin{eqnarray}
V(r) & = & -\frac{\alpha(T)}{r}e^{-r/r_{D}(T)},
\label{eq:VD0}
\end{eqnarray}
where $r_{D}$, Debye screening radius, is the decreasing function of temperature.  Roughly speaking, once the screening radius is smaller than the radius of $J/\psi$, the quarkonium will dissociate.  The dissociation of $J/\psi$ can be used as the signature of the formation of QGP as well as its temperature profile~\cite{sat2}.

One can calculate the heavy quark and antiquark potential using the AdS/CFT correspondence~\cite{mal,wit1}.  At zero temperature, the potential~\cite{jm}
\begin{eqnarray}
V=-\frac{4\pi^{2}\sqrt{\lambda_{4}}}{\Gamma(\frac{1}{4})^{4}}\left(\frac{1}{r}\right),
\label{eqnadspotential}
\end{eqnarray}
where $\lambda_{4}\equiv g^{2}_{YM}N \equiv 4\pi g_{s}N$ and $r$ is the distance between quark and antiquark.  One can see that the potential is proportional to $-1/r$.  However, the confining potential $\sigma r$ is dominant at large distance in cold nuclear matter.  Therefore we can see this model is more accurate to the deconfined phase where the $\sigma$ disappears, which requires the temperature to be higher than the deconfinement temperature $T_{c}$.  In order to take into account the temperature and thermal phenomena of the gauge plasma, one can consider a Schwarzschild-anti-de Sitter Type IIB supergravity compactification~\cite{wit2}.        

In the gravitational dual picture, the meson is obtained using the fundamental string whose two ends are connected with the probe brane.  The two ends of this string represent the quark and antiquark separately.  The position of the probe brane sitting at certain point in the radial direction is proportional to the mass of the quark.  Since one expect that the presence of the probe brane would not affect the geometry in the leading order, it is necessary to put the probe brane far away from the branes which produce the gravitational background, and hence the large mass of the quark and antiquark.  In the dual picture, there are two configurations which are competing with each other.  One is the fundamental string connecting the quark and antiquark which corresponds to the bound state.  The other is the two strings stretching from the probe brane to the horizon which represents the free quark and antiquark.  At given temperature, one needs to compare the energy difference of these two configurations.  It is used to determine the critical separation between quark pairs at which the bound states dissociate.  The dissociation of bound state really implies the situation when the energy of the parallel strings is lower than the connected configuration.  Then it is always preferred energetically for the pair to be separated as free quarks.  Using this strategy, one would see that at finite temperature the potential between heavy quark and antiquark exhibits short-range asymptotic behaviour~\cite{rty} which is qualitatively similar to the color-screened potential in Eqn.~(\ref{eq:VD0}). 
     
The results from the gravity dual theory show that the static heavy quark potential has $-1/r$ behaviour at zero temperature and short-range asymptotic behaviour at the finite temperature.  This results in scaling of the screening length $L^{*}$ with $1/T$ for states with zero angular momentum.  It was also demonstrated that the screening length of the quark antiquark state scales with velocity of the moving meson as $L^{*}\sim (1-v^{2})^{1/4}$~\cite{lrw}.  However, one may ask if we can obtain the effective potential for the excited heavy quark bound states with nonzero angular momentum from the gravity dual theory.  Excited heavy quark and antiquark bound states could have the non-zero angular momentum and spin.  The potential between heavy quarks which can move arbitrarily in the bound states would depend on the spin, angular momentum, as well as the velocity~(e.g. see explicit form calculated using dual QCD in Ref.~\cite{bbz}).  Therefore, the potential would be more complicated than the static potential.  Once the complete potential between heavy quark pairs is obtained, we can use it in the Schr\"{o}dinger equation to get the complete spectrum of the mesonic states~(see e.g. Ref.~\cite{qr}). 

Due to the success of the gravity dual theory to describe the static potential of heavy quark pair, we are optimistic and hope to estimate the equivalent potential for the excited bound states from the gravity dual picture.  To obtain the gravity dual picture, we boost the gravitational background metric~\cite{rot} for two different models: $AdS_{5}$ and Sakai-Sugimoto model.  Taking advantage of the numerical technique, we calculate the effective potential of excited bound state as a function of the separation between quark pairs $r$, at fixed angular momentum $J$ and temperature $T$.  The result shows that for each fixed angular momentum $J$ and given temperature $T$, the effective potential becomes positive at small distance due to the angular momentum effect being dominant in this region.  At the intermediate distance, the potential turns negative where the static binding potential dominates.  Finally it becomes positive again at large distance because of screening effect.  We also calculate the static potential for $\ell\equiv J =0$ at nonzero temperature for two different models.  The difference between the two models is discussed.

This paper is organized as the following.  In Section II, we demonstrate how introduction of the angular momentum barrier allows the effective potential to become positive at large distance, and therefore the existence of the ``screening length''.  In Section III, we analytically and numerically obtain the effective potential for $\ell=0$ and excited $\ell=1, 2$ bound states of heavy quark pair using the gravity dual models.  The shape of effective potential produced by the gravity dual models are qualitatively similar to the expected 4-dimensional result from Section II.  Significant differences between the two gravity dual models, $n=3,4$, are discussed.  In Section IV, fitting of the effective potential for excited $\ell=1,2$ states to the asymptotic form $V_{eff}=C_{n}-\alpha_{n}/r^{2/(n-2)}$ motivated from $\ell=0$ case is performed and shows exceptionally good fit.  We summarize our results in Section V. 

\section{Heavy Quark and Antiquark Potential}

Because of the larger mass of heavy quark e.g. charm and bottom, the bound states of heavy quark and antiquark can be described using the potential model.  One can expect that the heavy quark and antiquark bound state may be described by the solutions of the nonrelativistic Schr\"{o}dinger equation~\cite{qr,ms}
\begin{eqnarray}
-\frac{\hbar^{2}}{2\mu} \bigtriangledown^{2} \Psi(\textbf{r}) + [V(\textbf{r})-E] \Psi(\textbf{r})=0
\label{eqn3}
\end{eqnarray}  
where $\mu$ is the reduced mass of the quark-antiquark, with $\mu=m/2$ for $c\bar{c}, b\bar{b}$ quarkonium states.  For the central potential, the wave function $\Psi(\textbf{r})$ can be written as 
\begin{eqnarray}
\Psi(\textbf{r})= R(r)Y_{\ell m}(\theta,\phi),
\label{eqndec}
\end{eqnarray}
where $R(r)$ is the radial wave function and $Y_{\ell m}$ is the spherical harmonic function.  With this substitution, the radial wave function satisfies
\begin{eqnarray}
-\frac{\hbar^{2}}{2\mu} \left(\frac{d^{2}}{dr^{2}}+\frac{2}{r}\frac{d}{dr}\right)R(r)-\left[E- V(r)- \frac{\ell(\ell+1)\hbar^{2}}{2\mu r^{2}}\right] R(r) =0.
\label{eqnschro}
\end{eqnarray}

One can define the effective potential 
\begin{eqnarray}
V_{eff}(r,\ell)= V(r)+\frac{\ell(\ell+1)\hbar^{2}}{2\mu r^{2}},
\label{eqneffective}
\end{eqnarray}
where $\ell$ is the quantum number for the orbital angular momentum.  The term induced by orbital angular momentum $\ell(\ell+1)\hbar^{2}/2\mu r^{2}$ forms an angular momentum ``barrier'' $\sim 1/r^{2}$ dominating at small distance $r$.  This angular momentum barrier is crucial when we consider fundamental differences between $\ell=0$ and $\ell > 0$ quarkonium states when they are submerged into the QGP as we will see in the following.

\begin{figure}[tb]
\centering
\includegraphics[width=3.2in]{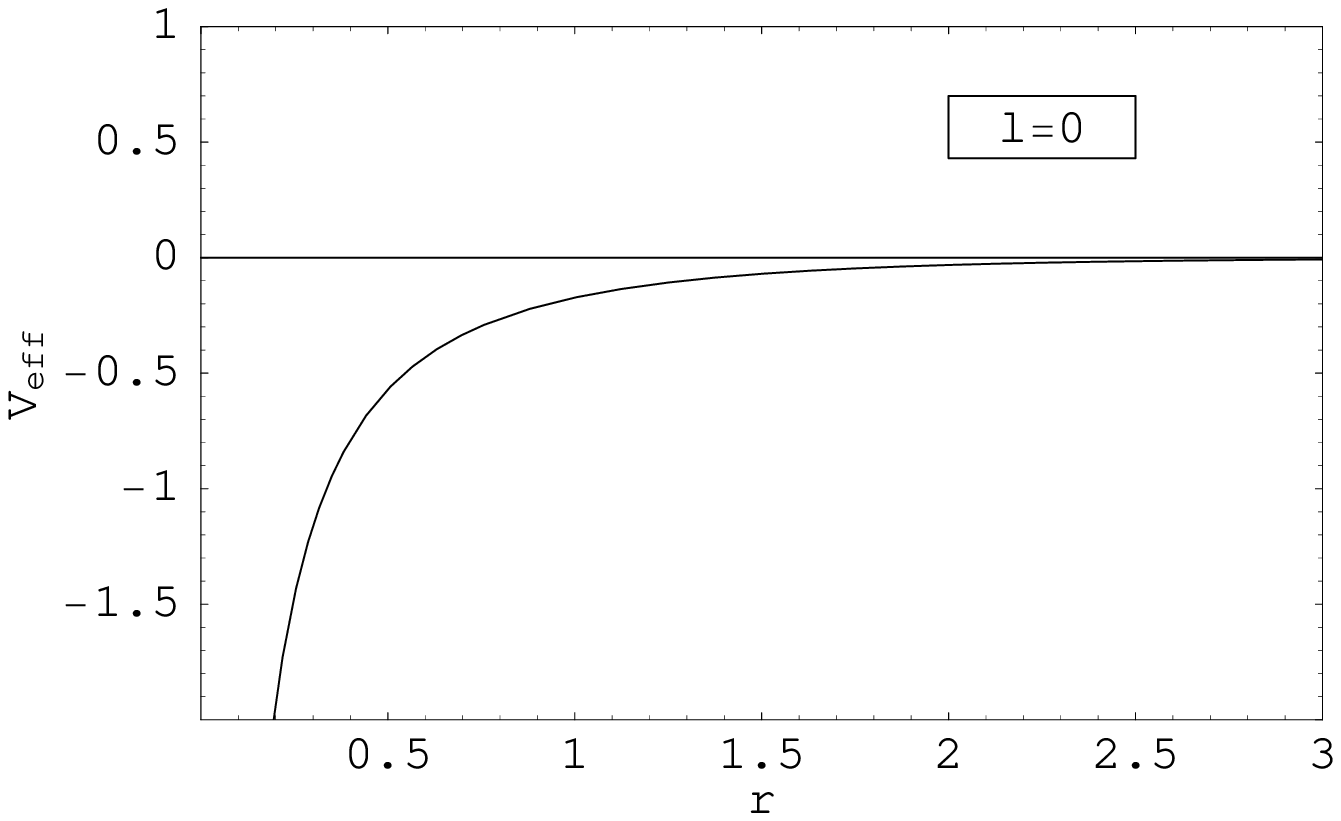}
\hfill
\includegraphics[width=3.2in]{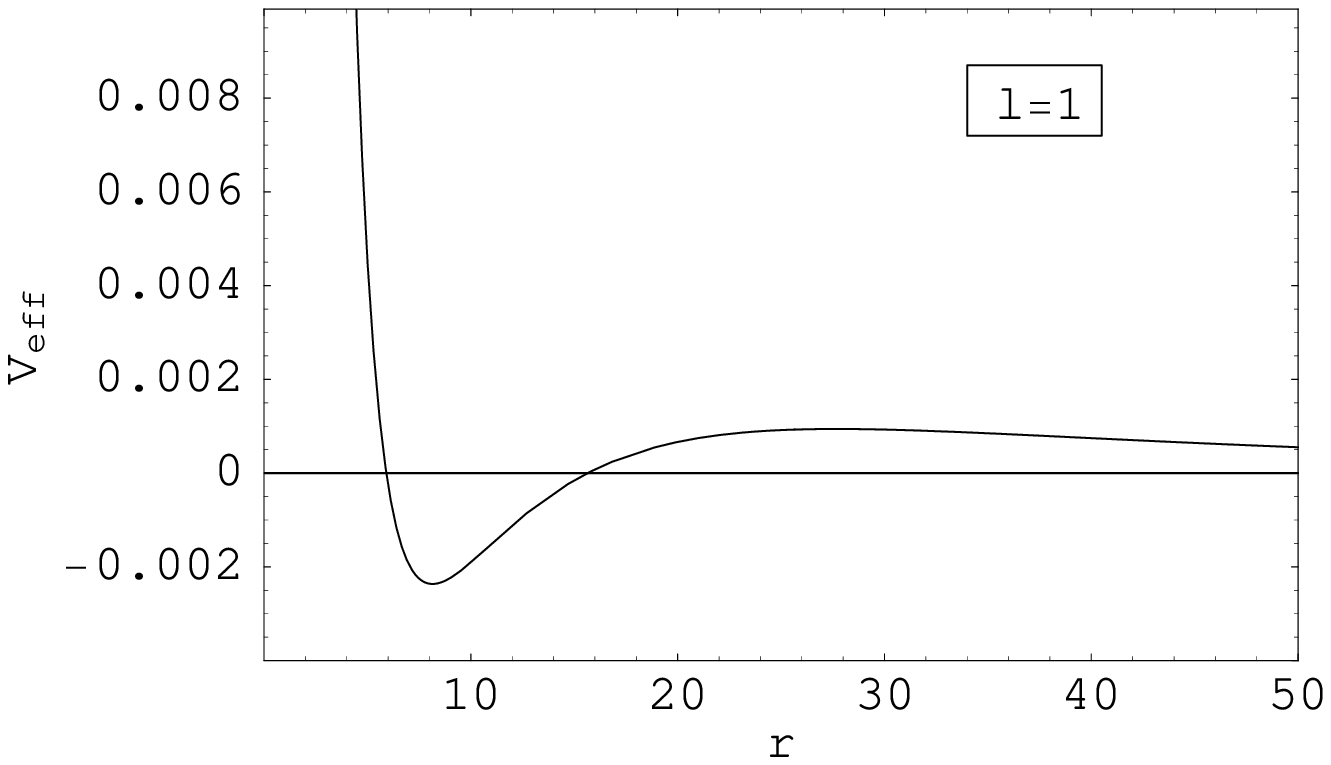}
\caption{The effective potential between quark and antiquark in the quark gluon plasma for $\ell=0$~(left) and $\ell=1$~(right) at finite temperature from Eqn.~(\ref{eq:V0}).  Potential barrier induced by the orbital angular momentum can modify the effective potential so that the potential actually becomes positive at distance larger than a finite ``screening length'' $L^{*}$ as shown in the figure on the right.  For demonstration purpose, we set $r_{D}=10$ for $\ell=1$ figure.  The screening length $L^{*}\sim 16$ is substantially larger than the screening radius $r_{D}=10$ in this case. }
\label{1-fig}
\end{figure}

In the deconfined phase such as QGP, the linear confining potential $\sigma r$ becomes zero.  The remaining is the Coulomb-like potential which would be screened by the plasma.  Analogous to the QED plasma, the screening QCD potential between quark and antiquark in the quarkonium submerged in the gauge plasma can be expressed as in Eqn.~(\ref{eq:VD0}).  The effective potential then becomes 
\begin{eqnarray}
V_{eff} & = & \frac{\ell(\ell+1)\hbar^{2}}{2\mu r^{2}}-\frac{\alpha(T)}{r}e^{-r/r_{D}(T)}
\label{eq:V0}
\end{eqnarray}
where $r_{D}$ is the Debye screening radius.  

The effective potential for $\ell=0$ state approaches zero from below but never actually crosses over and becomes positive.  On the other hand, the potential barrier induced when $\ell>0$ modifies the total potential such that at finite distance $L^{*}>r_{D}$, the effective potential changes sign and becomes positive for $r>L^{*}$ as is shown in Fig.~\ref{1-fig}.  To distinguish $L^{*}$ where $V_{eff}$ actually becomes zero from the Debye screening radius $r_{D}$, we will call it ``screening length'' hereafter.  

The ``melting'' or dissociation of the bound state of quark and antiquark could be understood as the tunneling of the quark~(antiquark) from the region where $V_{eff}<0$, through the region of potential barrier $V_{eff}>0$, to the large $r$ region where it is effectively free.  It is also possible that the thermal kinetic energy of the quark in the QGP~($\simeq k_{B}T$) is substantially larger than the binding effective potential and thus resulting in almost complete dissociation of the $\ell>0$ states~(see e.g. Fig.~\ref{1-fig}).  This could explain why the $\ell=0$ states are much harder to ``melt'' comparing to $\ell>0$ states.   

We can see that the angular momentum induces significantly distinctive features to the effective potential and consequently to the behaviour of the excited quarkonium states.  It is desirable to understand the more precise form of the potential as well as its dependence on the temperature so that we would gain more understanding of the properties of the QGP as well as the determining factors of the melting of quarkonia with varying angular momentum when they are submerged into the plasma.

\section{Effective Quark Antiquark Potential from Gravity Dual Models}

In Section II, one can see that nonrelativistic Schr\"{o}dinger equation may be used to solve the heavy quark bound states provided that we have the correct form of the potential.  The question is the exact nature of the screening potential between quark and antiquark in the deconfined phase at finite temperature.  Analogy to QED plasma suggests the form of the screening potential in the form of the Debye screening, Eqn.~(\ref{eq:VD0}).  This form is expected from the perturbative calculation~\cite{bn} but there is no guarantee it would remain valid in the regime where the 't~Hooft coupling is large.      

A complementary approach to calculate the effective potential when the gauge interaction is strongly coupled with the large 't~Hooft coupling is by means of the AdS/CFT correspondence~\cite{jm}.  One may ask if it is possible to obtain $\ell$ dependent effective potential $V_{eff}(r,\ell)$ in the gravitational dual picture.  In the following part, we are going to obtain this kind of effective potential by boosting the background metric.

In order to calculate the potential between the rotating quark and antiquark pair, we consider the following 5 dimensional metrics 
\begin{eqnarray}
ds^{2}=\left(\frac{u}{R_{n}}\right)^{n/2}\left(-f_{n}\left(u\right)dt^{2}+d\rho^{2}+\rho^{2}d\varphi^{2}+dz^{2} \right)+\left(\frac{R_{n}}{u}\right)^{n/2}\frac{du^{2}}{f_{n}\left(u\right)}  \label{eq:me1}
\end{eqnarray}
where
\begin{eqnarray}
f_{n}\left(u\right)=1-\frac{u_{h}^{n}}{u^{n}},  \label{eq:me2}
\end{eqnarray}
$u$ is the radial direction.  $z$ is the direction perpendicular to the plane of rotation and position of the horizon $u_{h}=\frac{16}{9}\pi^{2}R_{3}^{3}T^{2}, \pi R_{4}^{2}T$ for $n=3,4$ respectively.

The $n=3$ metric, known as Sakai-Sugimoto model~\cite{ss}~(we ignore details of the compact 5 dimensional manifold here assuming no Kaluza-Klein excitations with respect to these directions) in the high temperature phase~\cite{psz}, is the near-horizon limit of the metric induced by a configuration of $N_{c}$ D4-branes intersecting with $N_{f}$ D8-branes and $N_{f}$ anti-D8-branes.  The (anti-)D8-branes are located at $x_{4}=0,(L)$ of the compactified $x_{4}$ around which the D4-branes wrap.  The string theory in this background is dual to the maximally supersymmetric $SU(N_{c})$ Yang-Mills theory in $1+4$ dimensions with one dimension compactified on a circle with radius $R_{3}$.

The $n=4$ metric is the near-horizon limit of the metric induced by a configuration of $N$ D3-branes in the 1+9 dimensional background.  In this limit, the non-compact 5 dimensional subspace is approximately $AdS_{5}$.  The string theory in this background is dual to the supersymmetric $\mathcal{N}=4$ Yang-Mills in 1+3 dimensions when the other compact 5 dimensional manifold is $S_{5}$.
The horizon in the gravity picture induces the scale and Hawking temperature into the theory~\cite{wit2}.  These quantities are identified as $\Lambda_{QCD}$ and temperature of the quark gluon plasma in the dual gauge picture.

In order to calculate the potential between quark and antiquark in the gravity dual picture, it is suffice to consider the classical action of the string configuration.

The Nambu-Goto action is given by
\begin{eqnarray}
S=\frac{1}{2\pi}\int d\tau d\sigma\sqrt{det(G_{MN}\partial_{\alpha}X^{M}\partial_{\beta}X^{N})}.
\end{eqnarray}
From the above background metric in Eqn.~(\ref{eq:me1}),(\ref{eq:me2}), the generic action for a general worldsheet gauge, $\tau=t,~u=u(\sigma),~\rho=\rho(\sigma),~\varphi=\omega t$, is 
\begin{eqnarray}
S=\frac{1}{2\pi}\int dtd\sigma\sqrt{\left(\frac{u^{\prime2}}{f_{n}\left(u\right)}+\left(\frac{u}{R_{n}}\right)^{n}\rho^{\prime 2}\right)\left(f_{n}-\omega^{2}\rho^{2}\right)}, 
\end{eqnarray}
where $u^{\prime},\rho^{\prime}$ is the derivative with respect to $\sigma$.

In order to calculate the potential and solve the equation of motion,  we assign the following {\it ansatz} for the rotating connected-string configuration,
\begin{eqnarray}
t=\tau,\,\,\,\rho=\sigma,\,\,\, u=u\left(\rho\right),\,\,\,\varphi=\omega t.
\label{eq:g1}
\end{eqnarray}

The Nambu-Goto action becomes
\begin{eqnarray}
S=\frac{1}{2\pi}\int dtd\rho\sqrt{\left(\frac{u^{\prime2}}{f_{n}\left(u\right)}+\left(\frac{u}{R_{n}}\right)^{n}\right)\left(f_{n}-\omega^{2}\rho^{2}\right)},
\end{eqnarray}
where $u^{\prime}$ is the derivative with respect to $\rho$.

Then the potential for the connected string configuration whose ends located at $\rho=\rho_{0}$ is
\begin{eqnarray}
V=\frac{1}{\pi}\int_{0}^{\rho_{0}} d\rho\sqrt{\left(\frac{u^{\prime2}}{f_{n}\left(u\right)}+\left(\frac{u}{R_{n}}\right)^{n}\right)\left(f_{n}\left(u\right)-\omega^{2}\rho^{2}\right)}.
\label{eq:V}
\end{eqnarray}
The conserved angular momentum can be calculated from $\partial L/\partial \omega$ as
\begin{eqnarray}
J=\int_{0}^{\rho_{0}} d\rho\frac{\omega\rho^{2}\left(\frac{u^{\prime2}}{f_{n}\left(u\right)}+\left(\frac{u}{R_{n}}\right)^{n}\right)}{\sqrt{\left(\frac{u^{\prime2}}{f_{n}\left(u\right)}+\left(\frac{u}{R_{n}}\right)^{n}\right)\left(f_{n}\left(u\right)-\omega^{2}\rho^{2}\right)}}.
\label{eq:J}
\end{eqnarray}
We will identify the classical angular momentum $J$ of the classical string with the orbital angular momentum quantum number $\ell$ of the quarkonium~(instead of $\sqrt{\ell(\ell+1)}$) and use them interchangeably in this article.  The quantum mechanical spin of the quark and antiquark in the quarkonium will be ignored since we consider only the bosonic string in the gravity dual picture.     
 
The corresponding regulating potential for the parallel strings configuration can be calculated in another worldsheet gauge,
\begin{eqnarray}
t=\tau,\,\,\,u=\sigma,\,\,\, \rho=\rho_{0},\,\,\,\varphi=\omega t.
\label{eq:g2}
\end{eqnarray}

The potential then becomes
\begin{eqnarray}
V^{\prime}=\frac{1}{\pi}\int_{u_{c}}^{u_{max}}\sqrt{1-\frac{\rho_{0}^{2}\omega^{2}}{f_{n}(u)}} ~du
\label{eq:V1}
\end{eqnarray}
where $u_{c}=u_{h}/(1-\rho_{0}^{2}\omega^{2})^{1/n}$, the minimum distance in the $u$-direction for rotating parallel strings.  It is interesting to note that in the rotating metric, the position of horizon where parallel strings end is shifted from $u_{h}$ to $u_{c}=u_{h}/(1-\rho_{0}^{2}\omega^{2})^{1/n} > u_{h}$.  The equation of motion in this gauge for parallel strings configuration is trivially satisfied since $\partial L/\partial(\partial_{\sigma}u)=L$ for the Lagrangian $L$. 

On the other hand, the equation of motion for the connected-string configuration is
\begin{equation}
\frac{d}{d\rho}\left(\frac{1}{\sqrt{\left(\frac{u^{\prime2}}{f_{n}}+\left(\frac{u}{R_{n}}\right)^{n}\right)\left(f_{n}-\omega^{2}\rho^{2}\right)}}\frac{u^{\prime}}{f_{n}}\left(f_{n}-\omega^{2}\rho^{2}\right)\right)  \\\\\
-\frac{f_{n}^{\prime}\left(\frac{u^{\prime2}}{f_{n}^{2}}\omega^{2}\rho^{2}+\left(\frac{u}{R_{n}}\right)^{n}\right)+\frac{nu^{n-1}}{R_{n}^{n}}\left(f_{n}-\omega^{2}\rho^{2}\right)}{2\sqrt{\left(\frac{u^{\prime2}}{f_{n}}+\left(\frac{u}{R_{n}}\right)^{n}\right)\left(f_{n}-\omega^{2}\rho^{2}\right)}}=0.
\label{eom}
\end{equation}
For general situation where $\omega \neq 0$, we can numerically solve the equation of motion for a given boundary condition $u(\rho_{0})=u_{max},u^{\prime}(\rho_{0})\to \infty$ and given value of the $\omega$.  The relationship between $V_{eff}$ and other quantities such as $J$ and $r\equiv 2\rho_{0}$ can be obtained by eliminating the parameter $\omega$.

In this article, we set $u_{max}=20, R_{n}=1$ throughout our numerical analysis.  The $u_{max}$~(mass) dependence of $V_{eff}$ for $\ell=0$ case is given in subsection \ref{A}.  The dependence of the binding part of the potential on $R_{n}$, for $u_{max}\to \infty$ limit, is $R_{3}^3, R_{4}^2$ for $n=3,4$ respectively.  The choice of $R_{n}$ determines the strength of the binding potential of the quarkonium in the QGP and as we will see in Eqn.~(\ref{eq:LT}), $R_{4}=1$ leads to $\alpha_{4}\simeq 0.23$~(of the binding potential of the form $-\alpha_{4}/r$) which is close to the value used in conventional potential models~\cite{sat2}.       

\subsection{The potential for $\omega =0~(\ell=0)$ {\bf case}} \label{A}

For this case, we have additional constant of the motion since the Lagrangian does not depend on $\rho$ explicitly,
\begin{eqnarray}
A \equiv L-u^{\prime}\frac{\partial L}{\partial u^{\prime}} = \frac{f_{n}\left(\frac{u}{R_{n}}\right)^{n}}{\sqrt{u^{\prime2}+f_{n}\left(\frac{u}{R_{n}}\right)^{n}}},
\end{eqnarray}
from which
\begin{eqnarray}
r & = & 2\int_{u_{b}}^{u_{max}}\frac{R_{n}^{\frac{n}{2}}u_{0}^{\frac{n}{2}}}{\sqrt{\left(u^{n}-u_{h}^{n}\right)\left(u^{n}-u_{b}^{n}\right)}}\, du,
\label{eq:L0}
\end{eqnarray}
where $u_{b}=\left(u_{h}^{n}+u_{0}^{n}\right)^{\frac{1}{n}}, u_{0}^{n}=A^{2}R_{n}^{n}$, and $u_{max}$ is position of the probe brane.
The effective potential $V-V^{\prime}$, from Eqn.~(\ref{eq:V}) and Eqn.~(\ref{eq:V1}), can then be expressed as
\begin{eqnarray}
V-V^{\prime} & = & \frac{1}{\pi}\left(\int_{u_{b}}^{u_{max}}\frac{\sqrt{u^{n}-u_{h}^{n}}}{\sqrt{u^{n}-u_{b}^{n}}}\, du - \int_{u_{h}}^{u_{max}} ~du \right) .
\label{eq:Veff}
\end{eqnarray}
We will first consider the effective potential in $u_{max}\to \infty$ limit and then calculate the leading order $u_{max}$~(mass) dependence subsequently.

$\bullet$ \underline{{\bf Effective potential in} $u_{max}\to \infty$ {\bf limit}}: \\

For $T=0$ in the $u_{max}\to \infty$ limit of this $\omega=0~(\ell=0)$ case, the regulated potential becomes
\begin{eqnarray}
V_{eff}(r,T=0)& = & -\frac{1}{\sqrt{\pi}}\frac{\Gamma(1-\frac{1}{n})}{\Gamma(\frac{1}{2}-\frac{1}{n})} \left[ \frac{2\sqrt{\pi}}{nr}\frac{\Gamma(1-\frac{1}{n})}{\Gamma(\frac{3}{2}-\frac{1}{n})}R_{n}^{n/2} \right]^{1/(\frac{n}{2}-1)}.
\label{eq:Vt0}
\end{eqnarray}
The potential has the form $V\sim  -\lambda_{5}/r^{2},-\sqrt{\lambda_{4}}/r$ with the 't~Hooft coupling $\lambda_{5,4}=27\pi R_{3}^3/2, R_{4}^4$ for $n=3,4$ respectively, in agreement with Ref.~\cite{son}.   

For nonzero temperature in $u_{max}\to \infty$ limit, by using asymptotic expansion as in Ref.~\cite{rty}, we can obtain the leading order contribution with respect to $\lambda$ as~(see generic form for general $n$ in Appendix A) 
\begin{eqnarray}
V_{eff}(r,T)& = & C_{n}(T)-\alpha^{\prime}_{n}\frac{\lambda^{1/(n-2)}}{r^{2/(n-2)}}\left[1+O((rT)^{2n/(n-2)}) \right],
\label{eq:Vt}
\end{eqnarray}  
where $\lambda=\lambda_{5,4}$ for $n=3,4$ and $C_{n}(T)$ is a constant which is the increasing function with respect to the temperature and that $C_{n}(0)=0$,
\begin{eqnarray}
C_{n}(T)& = & \frac{u_{h}}{\pi} \\
        & = & \left(\frac{2}{3}\right)^{5}\lambda_{5}T^{2}, ~\sqrt{\lambda_{4}}T  \qquad \qquad ~\mbox{  for $n=3,~4$}
\end{eqnarray}
and
\begin{displaymath}
\alpha^{\prime}_{n} = \left\{
\begin{array}{l}
(\frac{32}{27\pi^{2}})(\Gamma(\frac{2}{3})\Gamma(\frac{1}{2})/\Gamma(\frac{1}{6}))^{3}\simeq 0.00963 \qquad \qquad \quad ~~\mbox{  for $n=3$},  \\ [5mm]
\frac{2}{\pi}(\Gamma(\frac{3}{4})\Gamma(\frac{1}{2})/\Gamma(\frac{1}{4}))^{2}\simeq 0.2285 \qquad \qquad \qquad \quad ~~~\mbox{  for $n=4$}.
\end{array}
\right.
\end{displaymath}

\begin{figure}[th]
\centering
\includegraphics[width=3.2in]{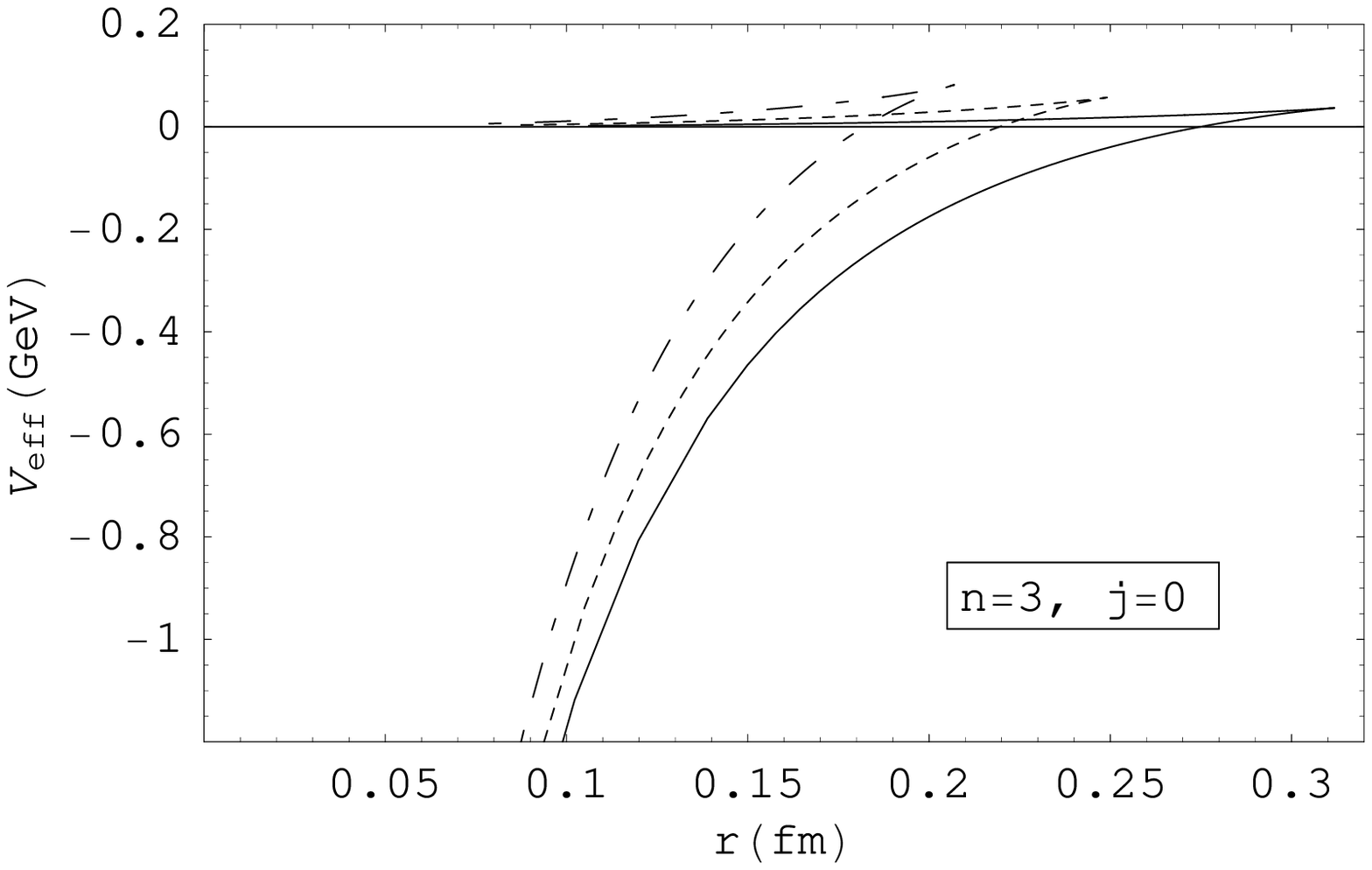}
\hfill
\includegraphics[width=3.2in]{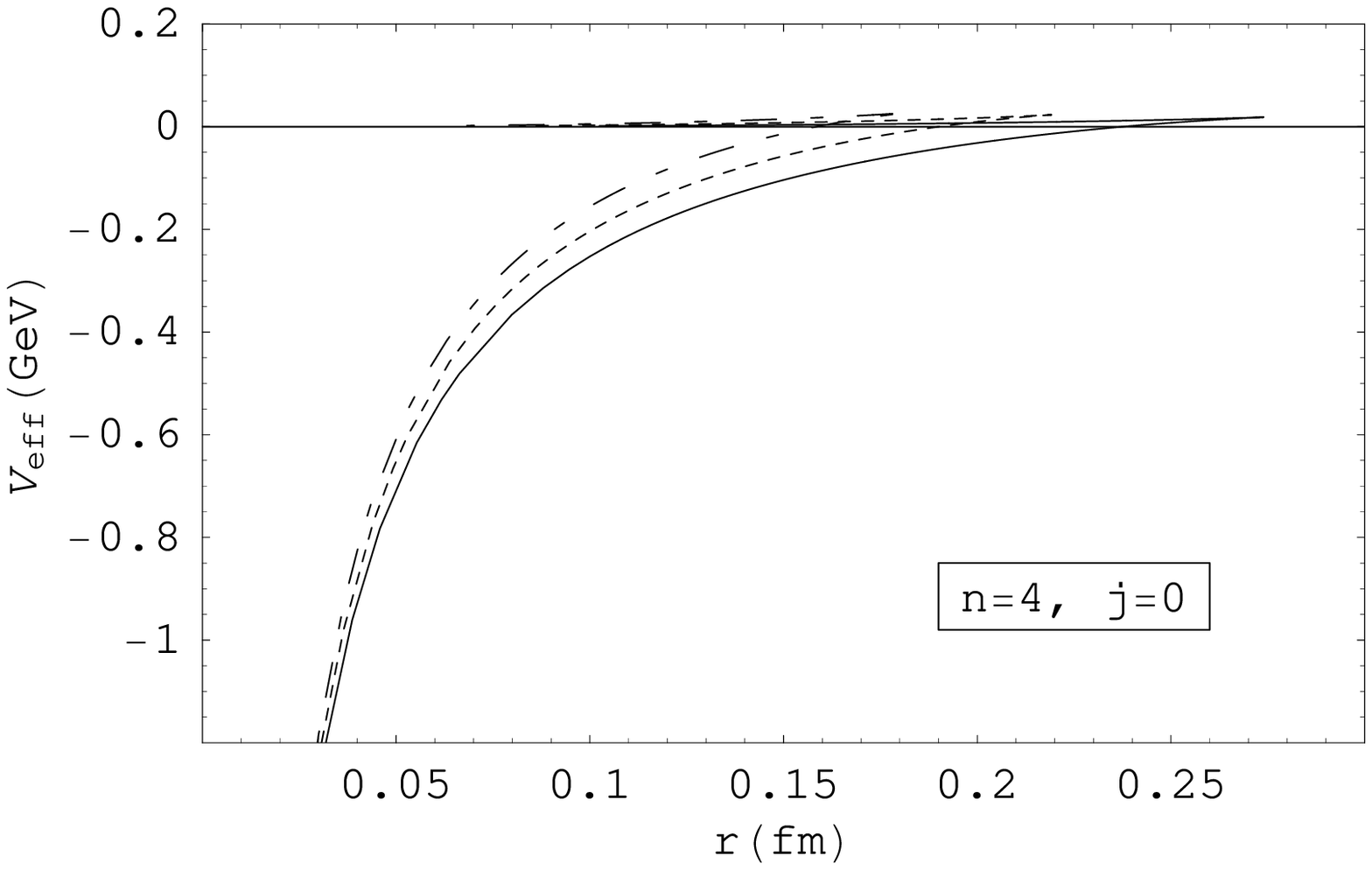}
\caption{The effective potential for $\ell\equiv J =0$ as a function of the distance between quark and antiquark for the metric $n=3$~(left) and $n=4$~(right) for the $\ell=0$ states.  The solid line is for $T=0.20$ GeV, the dashed line is for $T=0.25$ GeV, and the dashed-dot line is for $T=0.30$ GeV.}
\label{2-fig}
\end{figure}

The other terms of the potential are of the positive power of $rT$ and therefore suppressed as long as $r<1/T$.  From Eqn.~(\ref{eq:Vt}), it is obvious that $V_{eff}(r=L^{*},T)=0$ gives, at the leading order,
\begin{eqnarray}
L^{*}T & = & \sqrt{\alpha^{\prime}_{3}(3/2)^{5}}, ~\alpha^{\prime}_{4}\simeq 0.2704,~0.2285 \quad \quad \quad \mbox{for $n=3, 4$}. \label{eq:LT}
\end{eqnarray}
Note that the screening length $L^{*}$ is independent of $\lambda$ regardless of $n$.  

The numerical plots of the effective potential at nonzero temperature for each gravity dual are presented in Fig.~\ref{2-fig}.  We found that the fits to $1/r^{2},1/r$ for $n=3,4$ are valid with great accuracy for the physical range of $r$ close to the screening length into the binding region.  The asymptotic expansion works very well as long as the physical range of $r$ under consideration is not too large.  This provides a strong motivation to fit the form Eqn.~(\ref{eq:Vt}) to the binding region of more complicated cases with $\ell>0$ as we will see in Section IV. 

$\bullet$ \underline{$u_{max}$~{\bf (mass) dependence of the effective potential}}: \\

We have set $u_{max}\to \infty$ for the above analytical calculations.  The limit $u_{max}\to \infty$ provides universal form of potential for heavy quarkonia in the very large mass limit.  On the other hand, the specific mass dependence of the potential can be achieved when we fix $u_{max}~(\sim m$, mass of the quark) to some large finite value.

From Eqn.~(\ref{eq:Veff}), we can explicitly express $u_{max}$-dependent piece of $V_{eff}$ as
\begin{eqnarray}
V_{eff} & = & V_{eff}(u_{max}\to \infty)+V_{eff}(u_{max}) \\
V_{eff}(u_{max}) & = & -\frac{1}{\pi}\left(\int_{u_{max}}^{\infty}\frac{\sqrt{u^{n}-u_{h}^{n}}}{\sqrt{u^{n}-u_{b}^{n}}}\, du - \int_{u_{max}}^{\infty} ~du \right) \\
                 & = & -\frac{u_{max}}{\pi}\left( \int_{1}^{\infty}\frac{\sqrt{1-(h/y)^{n}}}{\sqrt{1-(b/y)^{n}}}\, dy - \int_{1}^{\infty} ~dy \right),
\end{eqnarray}
where $h,b\equiv u_{h,b}/u_{max}$.

By expanding with $A=h,b$,
\begin{eqnarray}
(1-(A/y)^{n})^{\pm1/2} & = & 1\mp \frac{1}{2}\left(\frac{A}{y}\right)^{n}+\frac{\pm\frac{1}{2}(\pm\frac{1}{2}-1)}{2!}\left(\frac{A}{y}\right)^{2n}+ ..., 
\end{eqnarray}
we obtain
\begin{eqnarray}
V_{eff}(u_{max}) & = & -\frac{u_{max}(b^{n}-h^{n})}{2\pi(n-1)}+\frac{u_{max}h^{n}b^{n}}{4\pi(2n-1)}+ ... ~.
\label{eq:Vumaxb}
\end{eqnarray}
We can eliminate $b$ by using Eqn.~(\ref{eq:L0}),
\begin{eqnarray}
r & = & 2\int_{u_{b}}^{u_{max}}\frac{R_{n}^{\frac{n}{2}}u_{0}^{\frac{n}{2}}}{\sqrt{\left(u^{n}-u_{h}^{n}\right)\left(u^{n}-u_{b}^{n}\right)}}\, du \\
  & = & 2R^{n/2}u_{max}^{(2-n)/2}\sqrt{b^{n}-h^{n}}\frac{b^{1-n}}{n}\left(\int_{0}^{1}-\int_{0}^{b^{n}}\right) \frac{y^{-1/n}}{\sqrt{1-y}\sqrt{1-(h/b)^{n}y}} dy \\
  & = & 2R^{n/2}u_{max}^{(2-n)/2}\frac{b^{(2-n)/2}}{n}\sqrt{1-\left(\frac{h}{b}\right)^{n}}\left[ a_{n}~{_{2}F_{1}}\left(\frac{n-1}{n},\frac{1}{2};\frac{3}{2}-\frac{1}{n};\left(\frac{h}{b}\right)^{n}\right)+O(u_{max}^{1-n})\right].
\end{eqnarray}
where $a_{n}$ is given in Appendix A.  Taking the leading contribution with respect to $O((h/b)^{n})$, and inverting to obtain
\begin{eqnarray}
b^{n} & = & \frac{1}{u_{max}^{n}}\left( \frac{2 a_{n}R^{n/2}}{nr} \right)^{2n/(n-2)} - \frac{n^{2}}{(n-2)(3n-2)}h^{n} \left[1+O((h/b)^{n}) \right]+...~.
\label{eq:bn}
\end{eqnarray}
We substitute Eqn.~(\ref{eq:bn}) into Eqn.~(\ref{eq:Vumaxb}) to obtain the leading order~(with respect to $u_{max},u_{h}$) dependence of effective potential on the mass~(defined at $T=0$), $m\equiv u_{max}/2\pi$ as
\begin{eqnarray}
V_{eff}(m)& \simeq & \frac{m^{1-n}}{(n-1)(2\pi)^{n}}\left[\frac{4(n-1)^{2}}{(n-2)(3n-2)}u_{h}^{n}-\left(\frac{2a_{n}R^{n/2}}{nr}\right)^{2n/(n-2)} \right].
\label{eq:Vumax}
\end{eqnarray}
The mass dependence contribution is sufficiently suppressed even at $u_{max}=20$.  For $n=3,4$, the leading order mass dependence $V_{eff}(m)\sim O(1/m^{2}),O(1/m^{3})$ respectively.  The $r$-dependence of $V_{eff}(m)$ is interestingly of order $O(1/r^{6}),O(1/r^{4})$ and the temperature dependence is $\sim T^{6},T^{4}$ for $n=3,4$ metric model.  It is interesting to compare this result with the mass-dependent potential calculated in other approaches such as pNRQCD~(see Ref.~\cite{bpsv} and references therein).  Note that our result, Eqn.~(\ref{eq:Vumax}), is applicable when the quarkonium is submerged in the QGP.
 
For $(\omega)~\ell>0$ case, since we are more familiar with the Coulomb-like potential $\sim 1/r$ from the $n=4$ $AdS_{5}$ gravity dual, we will first consider the effective potential calculated in this case in subsection \ref{B}.  The effective potential calculated in the $n=3$ Sakai-Sugimoto model will then be considered subsequently in subsection \ref{C}.  Comparison between the two gravity dual models will be discussed in subsection \ref{D}. 

\subsection{Effective Potential from $n=4$ metric, $AdS_{5}$ Model } \label{B}

From Eqn.~(\ref{eq:Vt0}), we can derive the well known result for the Coulomb-like potential at zero temperature for $\ell=0$ state, $V_{eff}\sim -\sqrt{\lambda_{4}}/r$.  At nonzero temperature, the asymptotic expansion also gives $V_{eff}\sim -\sqrt{\lambda_{4}}/r$ as the leading order contribution with respect to the large 't~Hooft coupling $\lambda_{4}$~\cite{rty}.  The next-to-leading order $C_{4}(T)$ term is a constant proportional to the temperature $T~(u_{h}(n=4)\sim T)$ and the other terms are of positive power of $rT$ and therefore suppressed as long as $r<1/T$.

As in Eqn.~(\ref{eq:V0}), we expect that inclusion of the angular momentum barrier brings in positive $1/r^{2}$ term dominant at small $r$ as well as reducing the value of the screening length $L^{*}$ when compared with $\ell=0$ case at the same temperature~(see Table I of Ref.~\cite{bl}).  The effective potential of the states with $\ell \equiv J=1,2$ at varying temperature is given in Fig.~\ref{3-fig}, \ref{4-fig} respectively.  

At $T=0, \ell=1$, the potential starts with positive value at small $r$ due to the angular momentum effect and turns negative around $r=0.04$ fm.  The effective potential becomes minimal at $r=0.06$ fm and becomes less negative at larger distance but never becomes positive.  It is obvious that as temperature gets higher, the potential becomes weaker and the screening length gets shorter.  Interestingly, the distance $r_{0}$ where the potential is minimum is the same regardless of the temperature.  Position of the minimum, $r_{0}$, is determined only by the angular momentum $J$ of the string.  As we will see later on, this interesting behaviour also exists in the case of $n=3$ Sakai-Sugimoto model for both $\ell=1,2$ cases~(as long as $T> T_{c}$).                

The behaviour of $\ell=2$ is very similar but somewhat fading out comparing to $\ell=1$ case.  This is the sign that the state with $\ell=2$ is less tightly bound and it is much easier to dissociate than $\ell=1$.  We can see that from Fig.~5 of Ref.~\cite{bl}, $\ell=2$ state melts around $T=0.24 $ GeV while $\ell=1$ does not melt as $T$ goes as high as $0.45$ GeV or so~(we will see later in Fig.~\ref{7-fig} that it melts around $T=0.59$ GeV)\footnotemark[1]\footnotetext[1]{ Even though the effective potential is negative in certain region, it is not necessarily true that it always admits quantum mechanical bound states in that region~(e.g. if the binding region is too shallow).  Therefore quarkonium, in practice, start to dissociate at lower temperature than the melting temperature determined classically when $V_{eff}\geq 0$~(for all $r$) as in Fig.~\ref{7-fig}.  We can think of the classically determined melting temperatures as the upper bounds on the melting temperatures above which complete dissociation is guaranteed.  }.  It should be emphasized that the position of minimum $r_{0}$ is fixed and independent of temperature as in $\ell=1$ case. 

The numerical results show that for given temperature $T$, one encounters two different regions for $r>r_{0}$ relevant to the melting of quarkonium.  When the separation between quark and antiquark $r < a_{J}/T$(where $a_{J}$ is the constant determining the screening length depending on $J$), there is the Coulomb-like behavior.  When $r > a_{J}/T$, one can see that the potential would cross over the $r$-axis and becomes positive.  The quark and antiquark become free and the bound state dissociates.

\subsection{Effective Potential from $n=3$ metric, Sakai-Sugimoto Model } \label{C}

At zero temperature for $\ell=0$ state, the potential from this metric is $V_{eff}\sim -\lambda_{5}/r^{2}$.  Even though this is not necessarily physical since we expect a phase transition to different metric when the temperature $T$ drops below $T_{c}$, this form of potential provides hint to what kind of $r$ dependence the effective potential has for generic $\ell$ at nonzero temperature.

At nonzero temperature, the asymptotic expansion for $\ell=0$ gives $V_{eff}\sim -\lambda/r^{2}+C_{3}(T)$.  The next-to-leading-order term $C_{3}$ is proportional to $T^{2}~(u_{h}(n=3)\sim T^{2})$ as opposed to $T$ of the $n=4$ case.  

For $\ell>0$, we present the results in Fig.~\ref{5-fig},~\ref{6-fig} for $\ell\equiv J= 1,2$ respectively.  The angular momentum barrier is dominant at small $r$.  The effective potential turns negative around $r=0.05\sim0.06$ fm for $\ell=1$, and around $r=0.085\sim0.095$ fm for $\ell=2$.  The minimum of potential is at $r_{0}=0.065~(0.11)$ fm for $\ell=1~(2)$.  The melting of $\ell=1,~2$ occurs at $T=0.50,~0.35$ GeV respectively as we can see in Fig.~\ref{7-fig}. 

The general shape of the potential curve for $\ell=2$ does not change much from $\ell=1$ in this metric model in contrast to the case of $n=4$ metric.  This is due to the smaller curvature with respect to the radial direction $u$ of Sakai-Sugimoto model.  The state with higher angular momentum will be rotating at larger distance in $u$ coordinate and the effect of curvature will be seen more distinctively between the two gravity dual models.  We can see that the melting temperature for $\ell=2$ states in $n=3$ model is around $0.35$ GeV, considerably higher than that of $n=4$, around $0.24$ GeV, as is shown in Fig.~\ref{7-fig}.  The $\ell=2$ states of $n=4$ metric model dissociate much earlier and easier.

From Fig.~\ref{5-fig},\ref{6-fig}, we can see that the screening length of this metric model is interestingly shorter than that of $n=4$ model for $\ell=1$, and larger than the value of $n=4$ for $\ell=2$ at the same temperature.  

\subsection{Comparison between the Two Metric Models} \label{D}

The curvature of the dual metric relevant to the physics of the gauge plasma is the one with respect to the radial direction $u$.  In our setup, the higher $n$ naturally implies higher curvature and we expect the $n=4, AdS_{5}$ model to show certain enhancing effects when compared to the $n=3$ Sakai-Sugimoto model.  The rotating string with higher angular momentum will be spinning at further distance from the horizon and we expect to see effects of curvature more distinctive than states with lower angular momentum.  

\begin{figure}[h]
\centering
\includegraphics[width=3.2in]{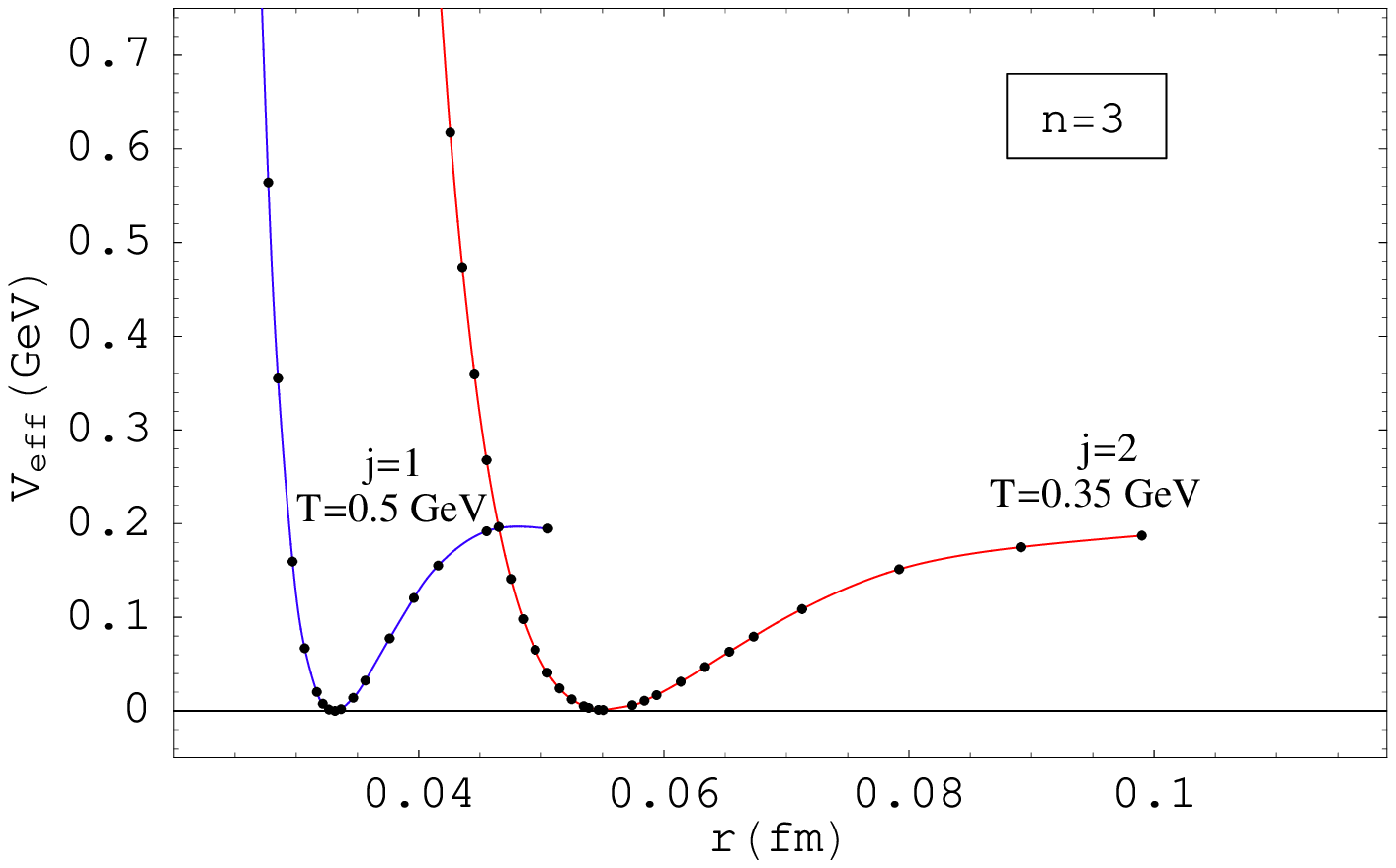}
\hfill
\includegraphics[width=3.2in]{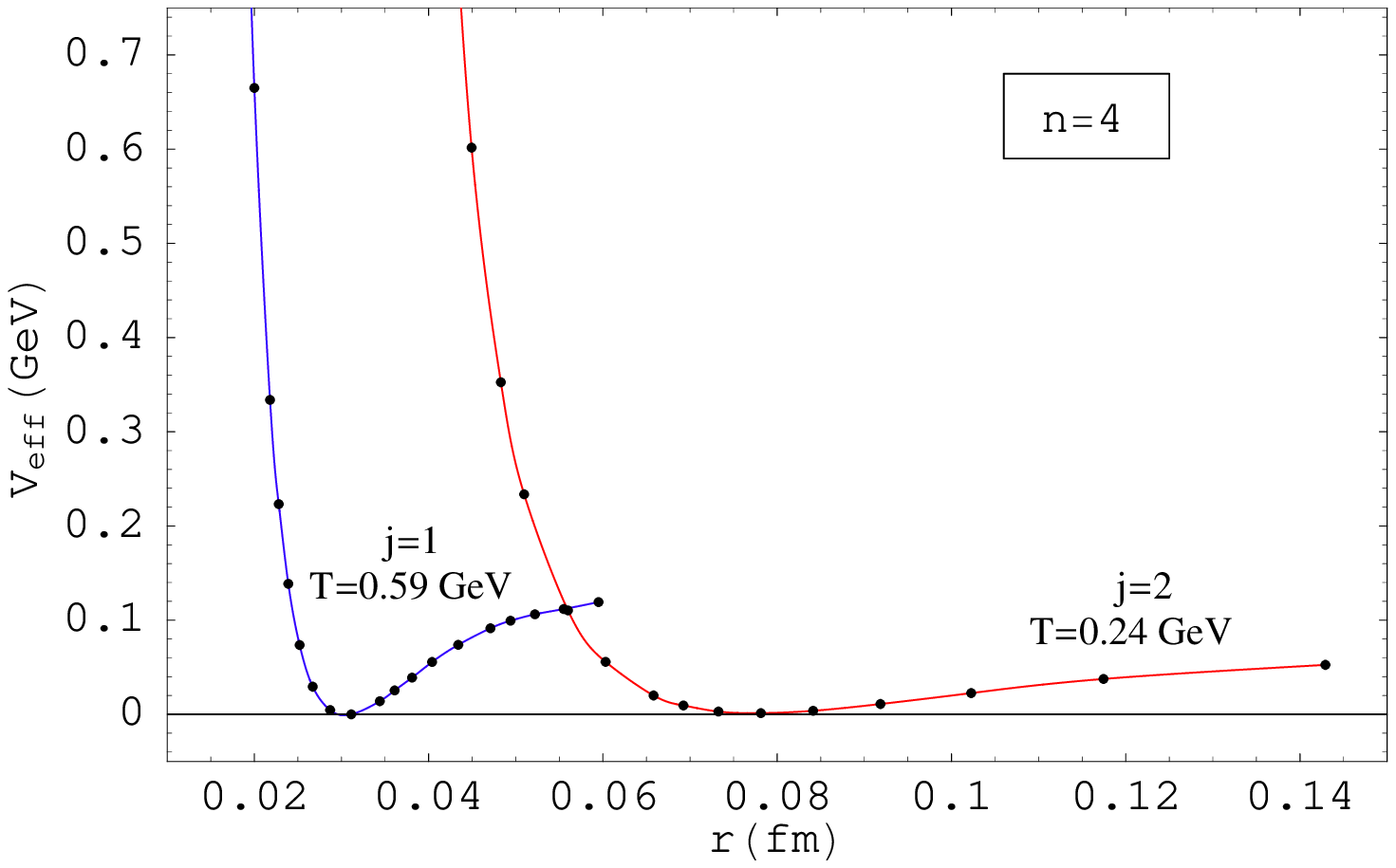}
\caption{The effective potential when the quarkonium melts, $n=3$, $\ell \equiv J=1, T=0.50$ GeV and $\ell \equiv J=2, T=0.35$ GeV on the left and $n=4$, $\ell \equiv J=1, T=0.59$ GeV and $\ell \equiv J=2, T=0.24$ GeV on the right. }
\label{7-fig}
\end{figure}

\begin{figure}[h]
\centering
\includegraphics[width=3.2in]{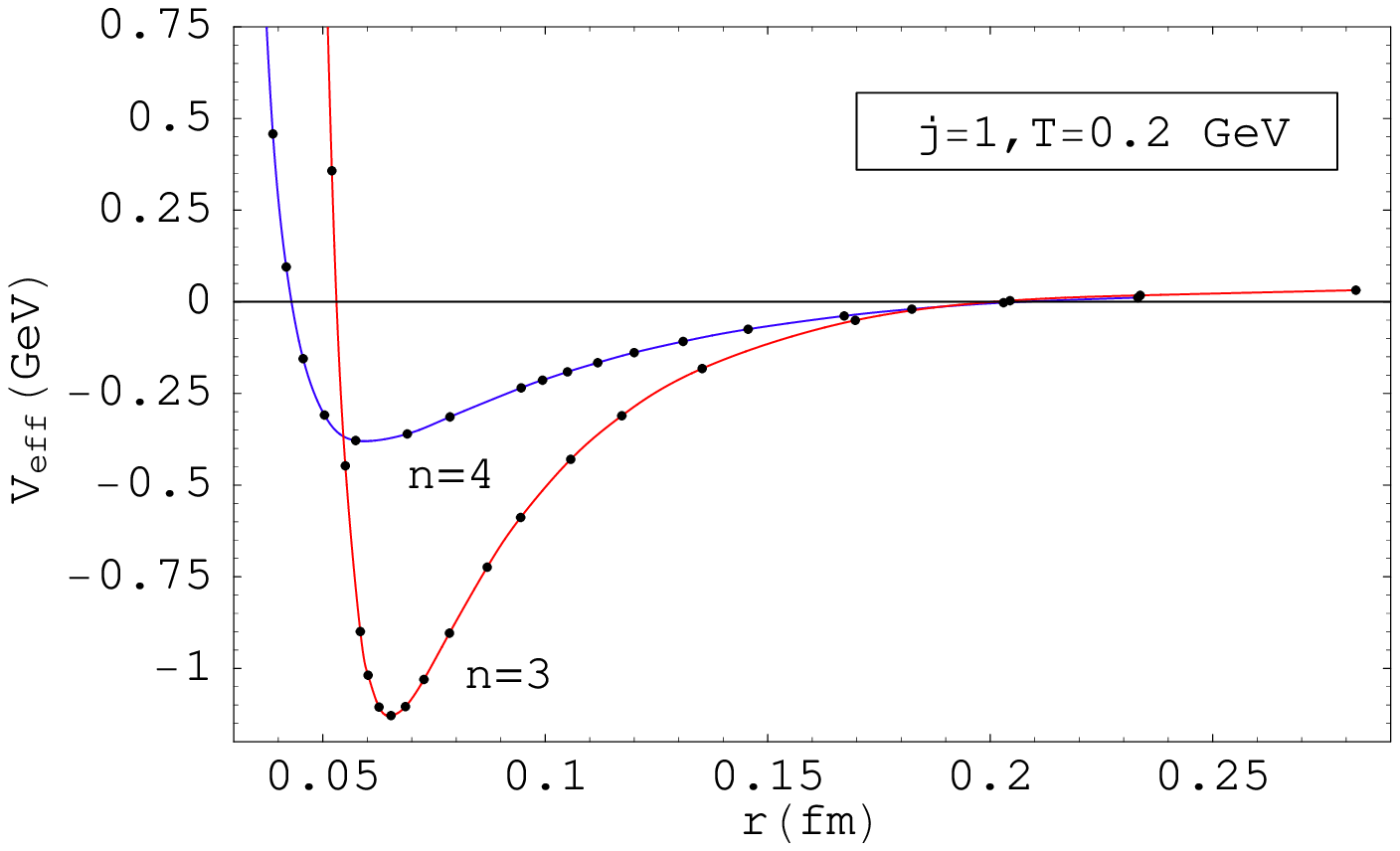}
\hfill
\includegraphics[width=3.2in]{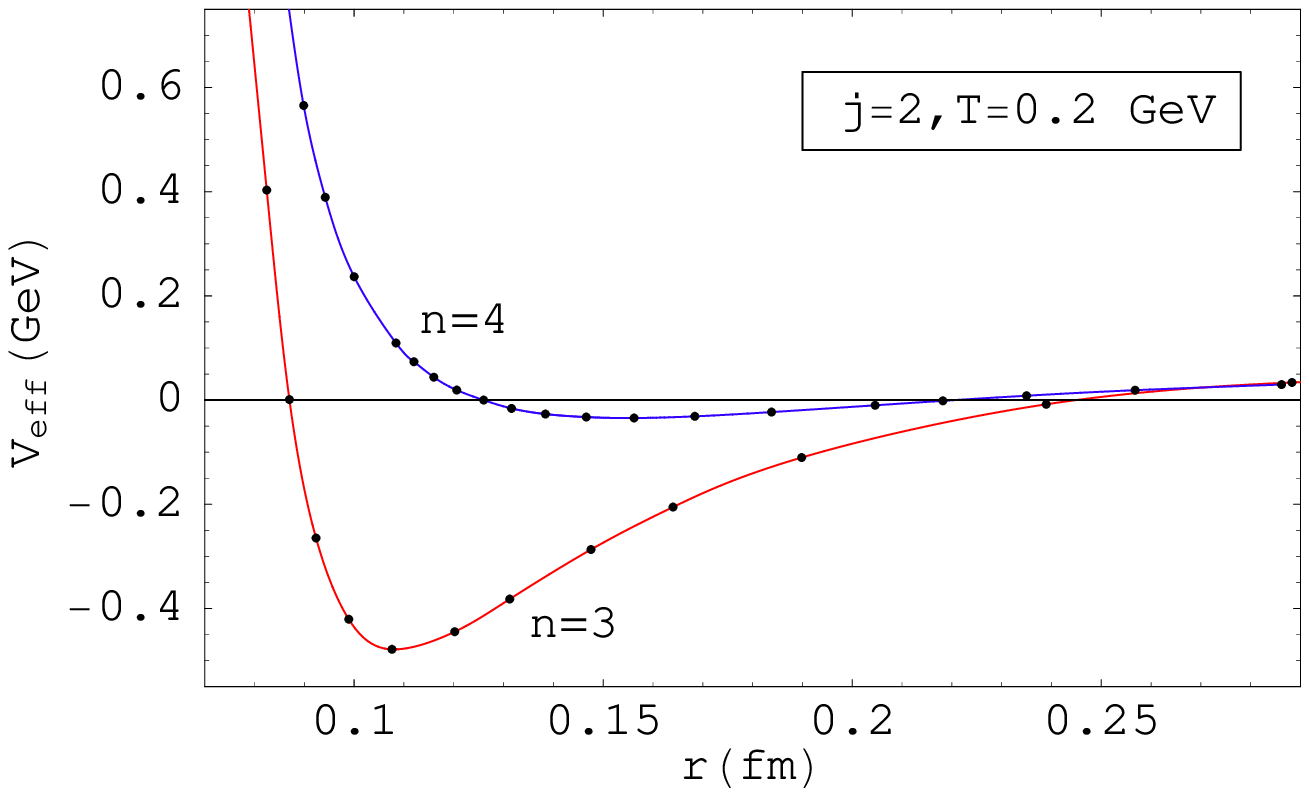}
\caption{$T=0.20$ GeV, $\ell \equiv J=1~(2)$ for $n=3,~4$ on the left~(right). }
\label{8-fig}
\end{figure}

\begin{figure}[h]
\centering
\includegraphics[width=3.2in]{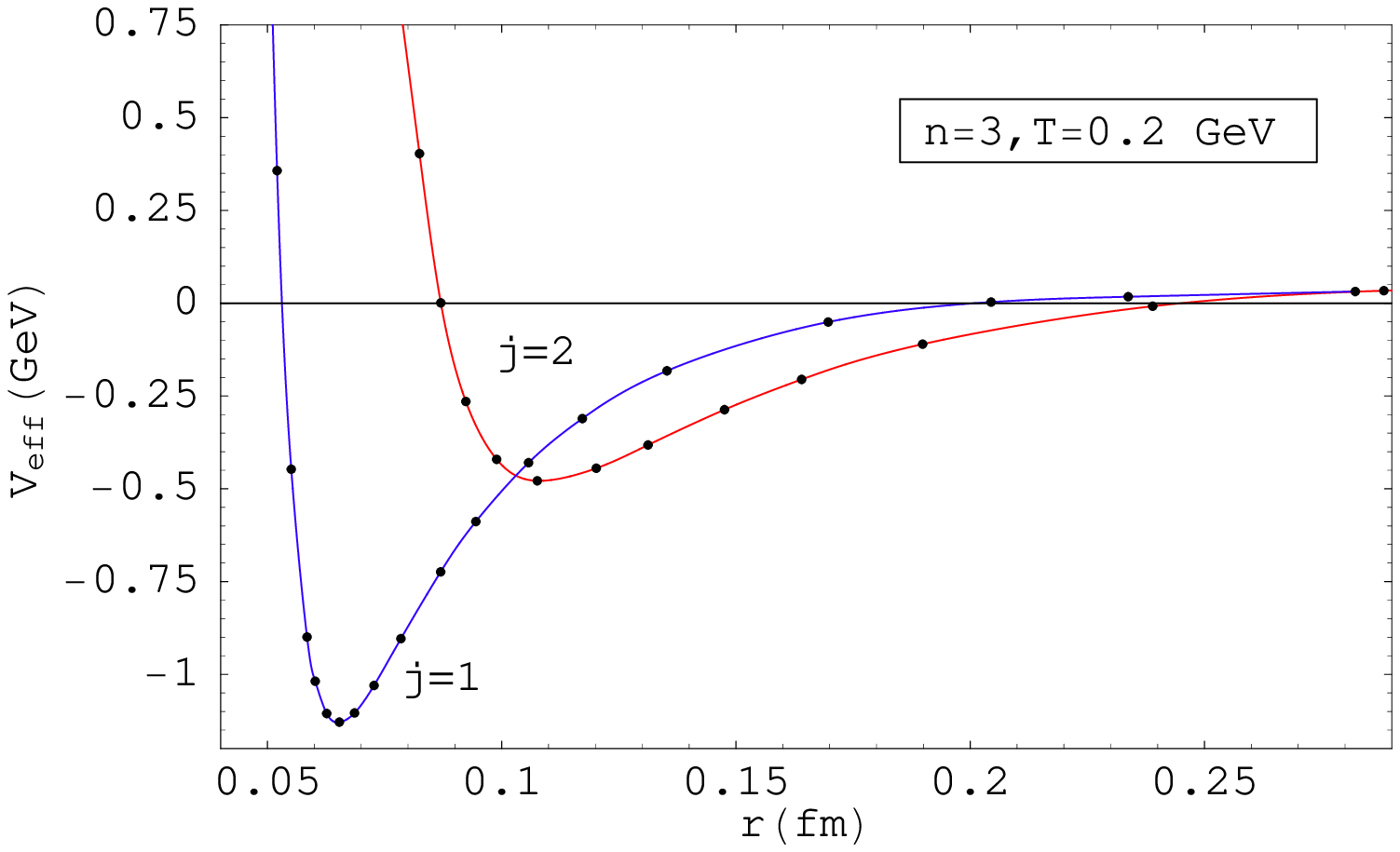}
\hfill
\includegraphics[width=3.2in]{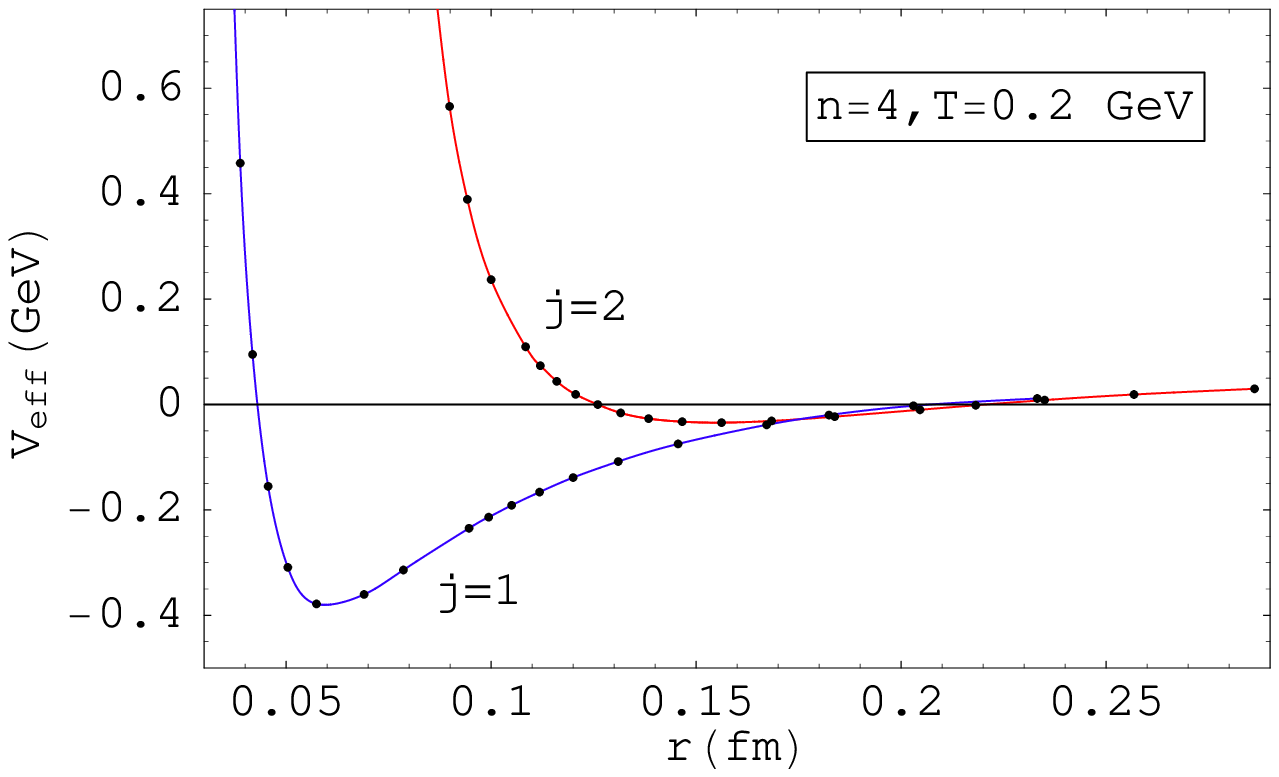}
\caption{$T=0.20$ GeV, $\ell \equiv J=1, 2$ for $n=3,~(4)$ on the left~(right). }
\label{9-fig}
\end{figure}

These curvature effects can be seen in Fig.~\ref{7-fig}-\ref{9-fig}.  Fig.~\ref{7-fig} shows the melting temperatures for $\ell=1,2$ states for each gravity dual model.  While the melting temperature for $\ell=1$ state of $n=4$ is higher than that of $n=3$, the melting temperature for $\ell=2$ state becomes drastically lower than that of $n=3$ model.

In Fig.~\ref{8-fig}, it is obvious that $n=3$ potential binds stronger than the potential of $n=4$ metric model, with much higher binding energies.  We can see that even at $T=0.20$ GeV, the $\ell=2$ state of $n=4$ already almost melts away while the same state of $n=3$ is still strongly bound.

Fig.~\ref{9-fig} represents comparison between states with $\ell=1,2$ for each metric.  The separation of $r_{0}$ in $n=3$ model is about $0.045$ fm, much lower than in $n=4$ where the gap of $r_{0}$ between $\ell=1,2$ is around $0.09$ fm.  The values of the minimum of potential $r_{0}$ are summarized in Table~\ref{t1}. 

\begin{table}[tb]
{\tabcolsep=.5cm  
\def\arraystretch{1.5}  
\medskip
\centering
\begin{tabular}{|c|c|c|} \hline
$n $&  $J=1$  &  $2$  \\ \hline
$3$ &  $0.065$ fm & $0.11$ fm  \\
$4$ &  $0.06$ fm & $0.15$ fm  \\ \hline
\end{tabular}  }
\caption[]
{The values of distance $r_{0}$ where the effective potential $V_{eff}(r,T)$ is minimum for the dual metric $n=3, 4$. } \label{t1}
\end{table}

It is observed in subsection \ref{C} that the screening length for $\ell=1~(2)$ from $n=3$ model is smaller~(larger) than the corresponding value from $n=4$ model.  Interestingly, from Eqn.~(\ref{eq:LT}) and Fig.~\ref{2-fig}, the screening length of $\ell=0$ state from $n=3$ metric model is also found to be larger than the value from $n=4$ model.

Finally, we also found numerically that the angular momentum $J$ does NOT depend on the temperature $T$, only depends on $\omega, r$.  We can see from Fig.~\ref{3-fig}-\ref{6-fig} that for fixed value of $J$ and $\omega$, $r$ is determined to be the same for each temperature curve~(points with the same $\omega$ are on the same $r$ once $J$ is fixed).  It appears that submerging quarkonium into the QGP at varying $T$ does not affect its angular momentum.  In other words, the interaction between QGP and quark~(antiquark) is radial and preserving angular momentum.   

\section{Fitting of the Effective Potentials in the Binding Region }

From Eqn.~(\ref{eq:Vt}) in the case of $\ell=0$, it is natural to wonder if this form of effective potential also describes the binding region of the more general $\ell>0$ cases.  By considering the shape of $V_{eff}$ in both metric models, a simple guess is that far away from the angular momentum barrier $\sim 1/r^{2}$, the binding region $r_{0}<r<L^{*}$ should be approximated to a good precision by the asymptotic form $V_{eff}=C_{n}(T)-\alpha_{n}(T)/r^{2/(n-2)}$.  We actually found that this is the case.

\begin{figure}[h]
\centering
\includegraphics[width=3.2in]{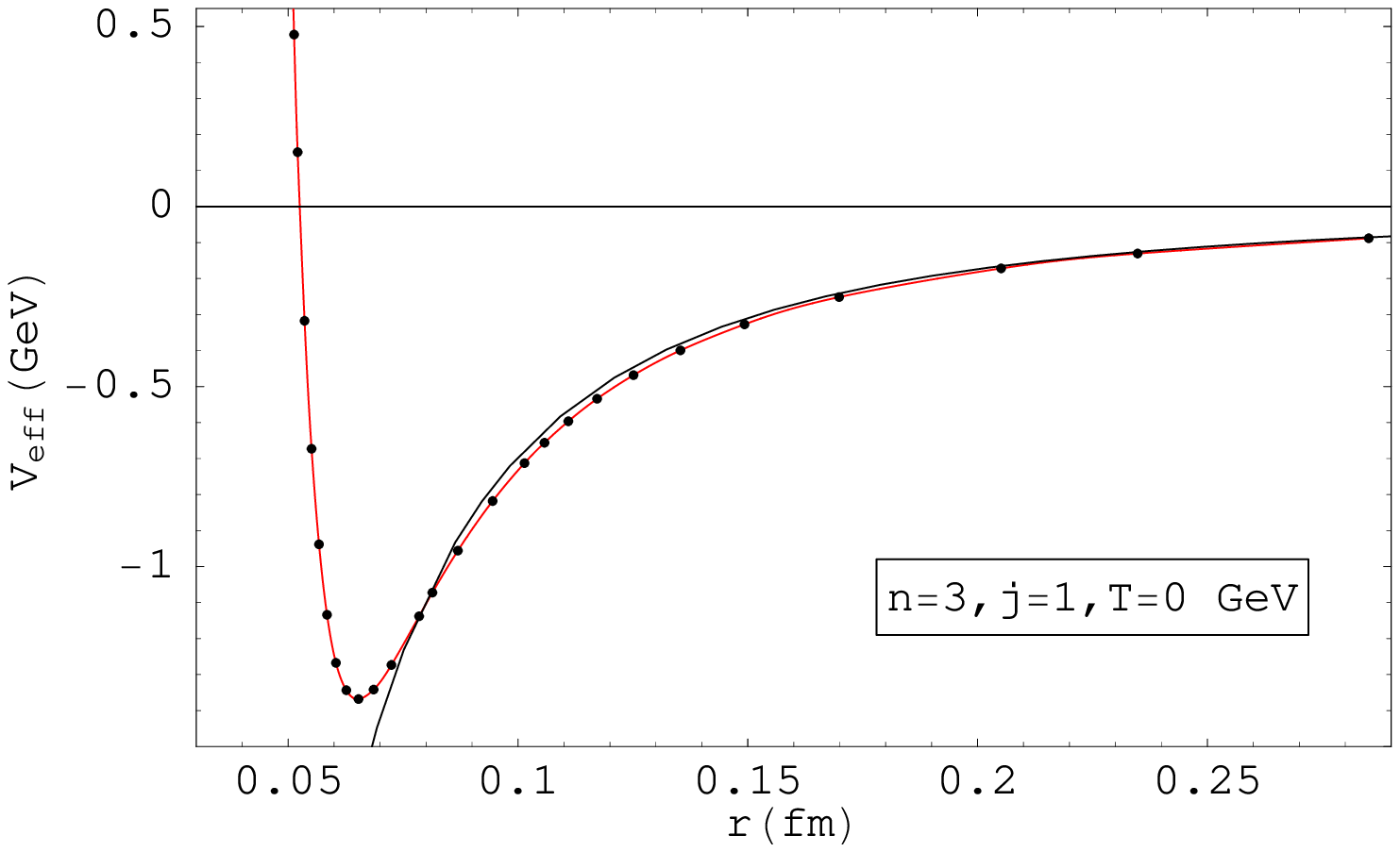}
\hfill
\includegraphics[width=3.2in]{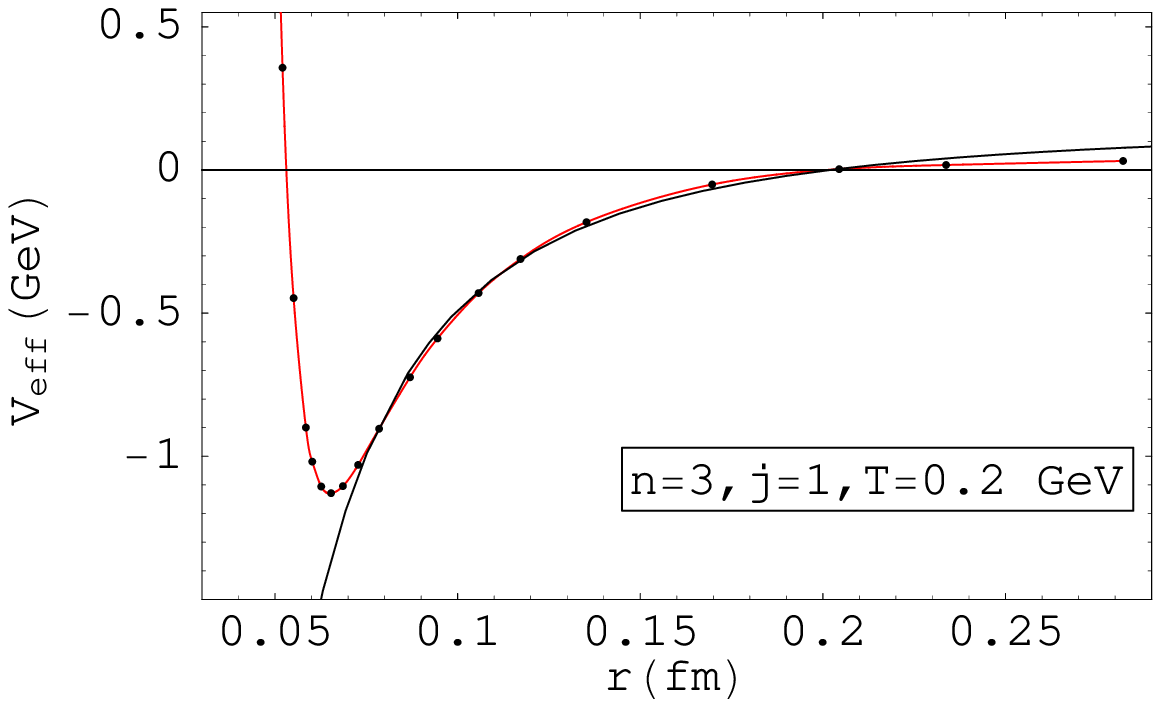}
\caption{The fitting of the physical binding region $r_{0}<r<L^{*}$ with $V_{eff}(r,T)=C_{3}-\alpha_{3}/r^{2}$ for $\ell\equiv J=1,~n=3$ dual metric. }
\label{10-fig}
\end{figure}

\begin{figure}[h]
\centering
\includegraphics[width=3.2in]{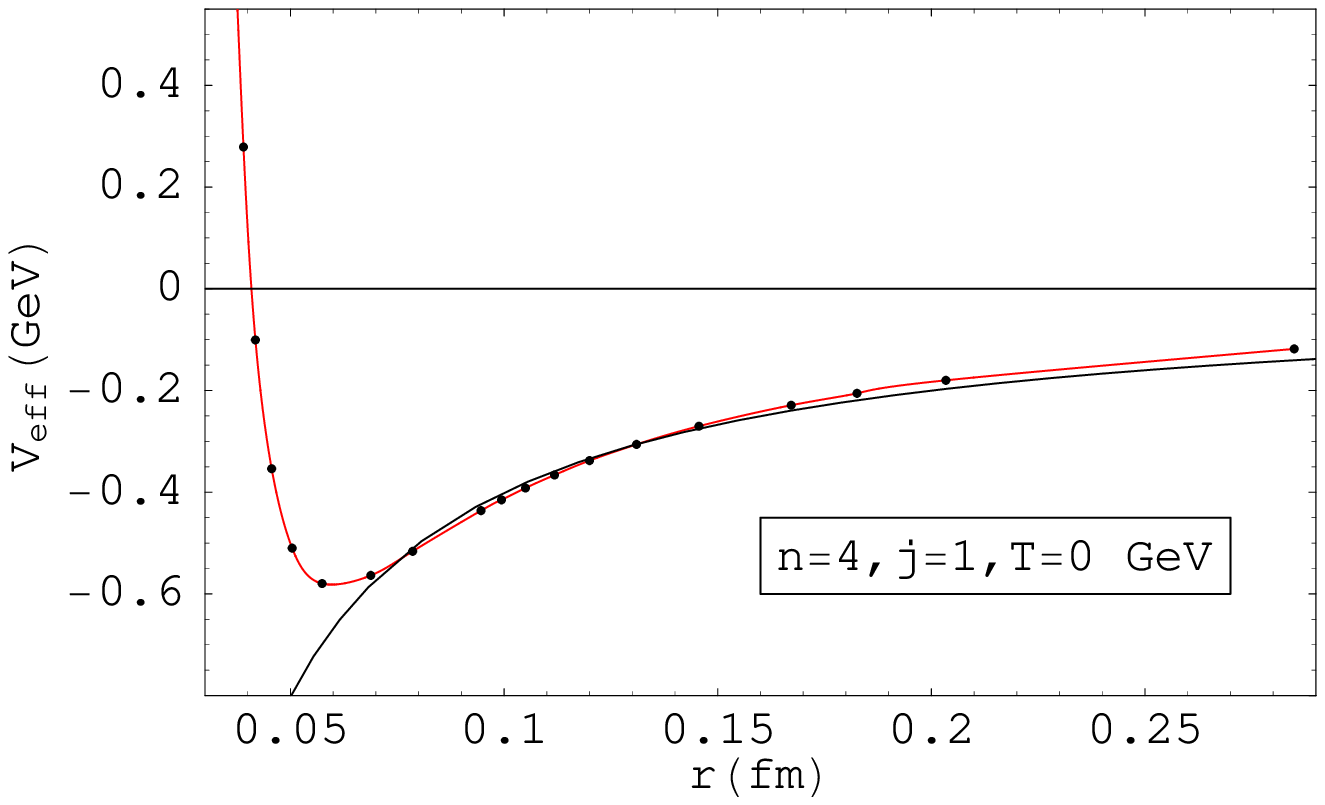}
\hfill
\includegraphics[width=3.2in]{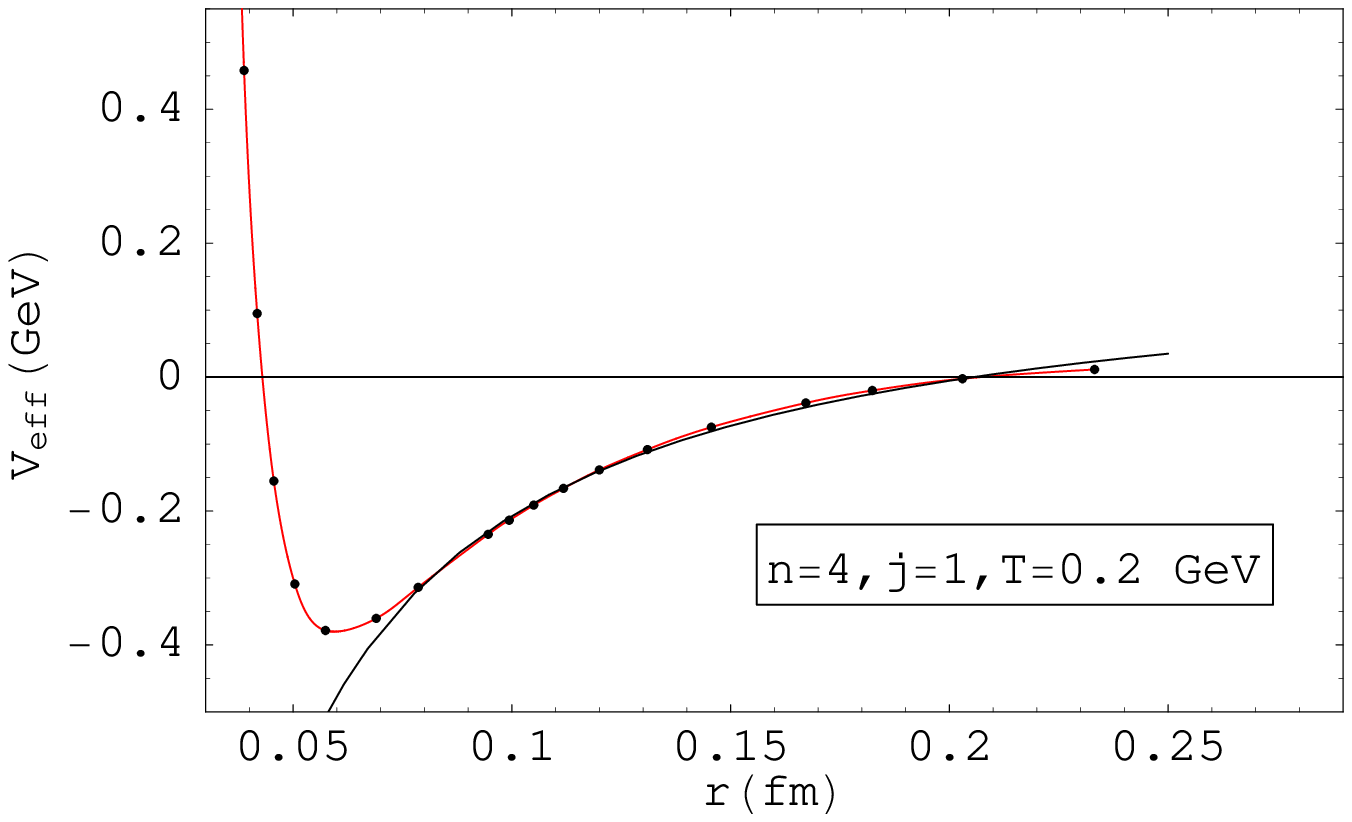}
\caption{The fitting of the physical binding region $r_{0}<r<L^{*}$ with $V_{eff}(r,T)=C_{4}-\alpha_{4}/r$ for $\ell\equiv J=1,~n=4$ dual metric. }
\label{11-fig}
\end{figure}

In Fig.~\ref{10-fig} and \ref{11-fig}, the fits to $\ell\equiv J=1$ for both metric models are demonstrated at $T=0,~0.20$ GeV.  We also found that the same fitting works very well for the case $\ell=2$.  Remarkably, the fitting works well at any temperature for both $\ell=1,2$ for both metric models.  The values of the best fit parameters for $\ell=1,~2$ are summarized in Table~\ref{t2},~\ref{t3} respectively.           

\begin{table}[h]
{\tabcolsep=.5cm  
\def\arraystretch{1.5}  
\medskip
\centering
\begin{tabular}{|c|c|c|c|c|} \hline
$T/$ GeV&  $\alpha_{3}/$ (GeV)(fm)$^{2}$  &  $\alpha_{4}/$ (GeV)(fm) &$C_{3}/$ GeV  &$C_{4}/$ GeV  \\ \hline
$0$ & $0.00695$ & $0.04004$ & $0$ & $0$ \\
$0.10$ & $0.00709$ & $0.04126$ & $0.0480$ & $0.11230$ \\
$0.20$ & $0.00648$ & $0.04037$ & $0.1591$ & $0.19646$ \\ 
$0.25$ & $0.00618$ & $0.0390$ & $0.2428$ & $0.23014$ \\
$0.30$ & $0.00596$ & $0.03758$ & $0.3484$ & $0.26211$ \\
$0.35$ & $0.00538$ & $0.03503$ & $0.4315$ & $0.28031$ \\ 
$0.40$ & $0.00447$ & $0.03194$ & $0.4802$ & $0.29134$ \\ \hline
\end{tabular}  }
\caption[]
{The values of the fitting parameters $\alpha_{n},~C_{n}$ for the binding region $r_{0}<r<L^{*}$ for the dual metric $n=3,~4$ for $\ell\equiv J=1$. } 
\label{t2}
\end{table}

\begin{table}[h]
{\tabcolsep=.5cm  
\def\arraystretch{1.5}  
\medskip
\centering
\begin{tabular}{|c|c|c|c|c|} \hline
$T/$ GeV&  $\alpha_{3}/$ (GeV)(fm)$^{2}$  &  $\alpha_{4}/$ (GeV)(fm) &$C_{3}/$ GeV  &$C_{4}/$ GeV  \\ \hline
$0$ & $0.01149$ & $0.04266$ & $0$ & $0$ \\
$0.10$ & $0.01135$ & $0.03779$ & $0.0450$ & $0.07936$ \\ 
$0.17$ & $0.01004$ & $0.03132$ & $0.1149$ & $0.11608$ \\
$0.20$ & $0.00926$ & $0.02639$ & $0.1465$ & $0.11995$ \\
$0.25$ & $0.00830$ & N/A & $0.2226$ & N/A \\
$0.30$ & $0.00623$ & N/A & $0.2631$ & N/A \\ \hline
\end{tabular}  }
\caption[]
{The values of the fitting parameters $\alpha_{n},~C_{n}$ for the binding region $r_{0}<r<L^{*}$ for the dual metric $n=3,~4$ for $\ell\equiv J=2$. } 
\label{t3}
\end{table}

\begin{figure}[h]
\centering
\includegraphics[width=3.2in]{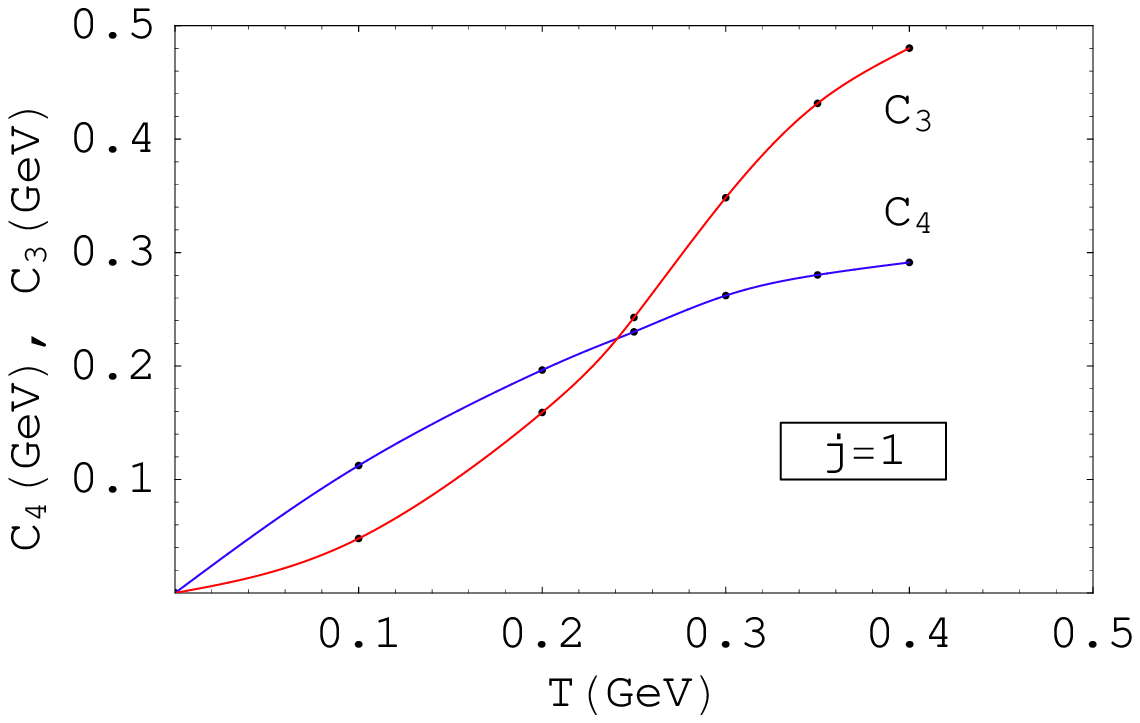}
\hfill
\includegraphics[width=3.2in]{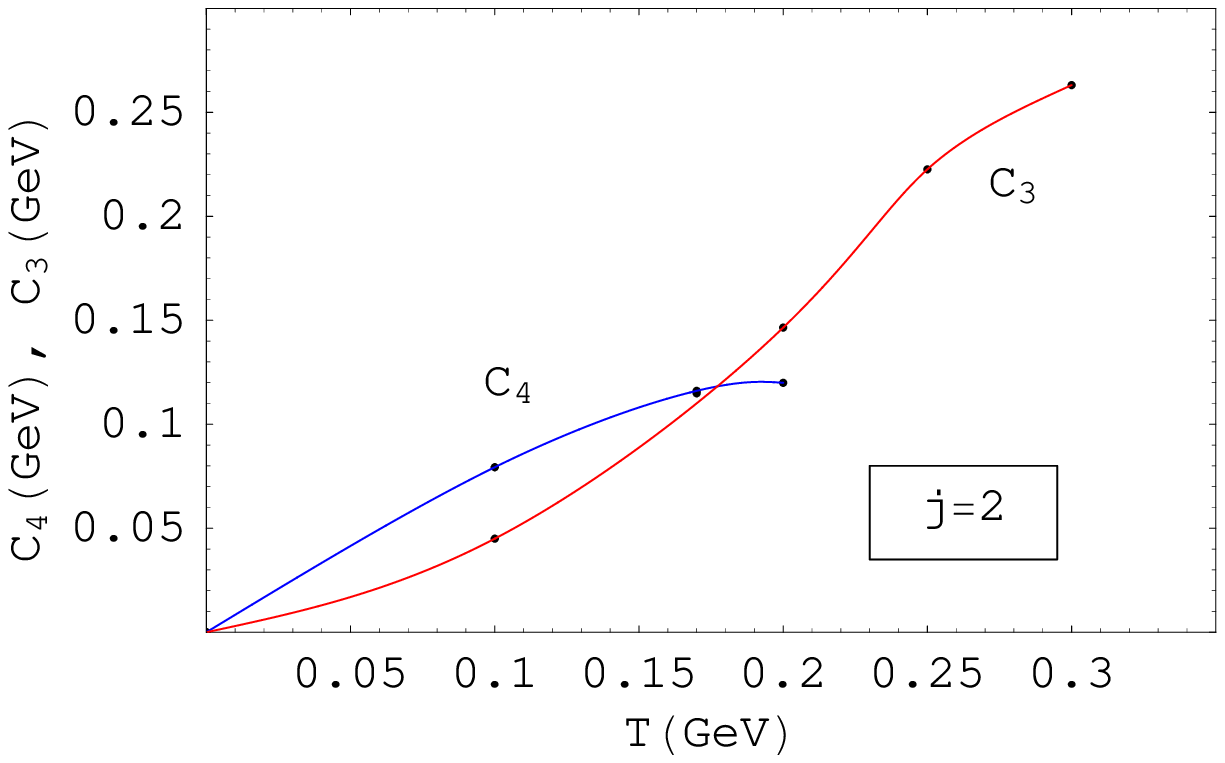}
\caption{The temperature dependence of the fitting parameter $C_{n}(T)$ for $\ell \equiv J=1$~(left),~$2$~(right). }
\label{12-fig}
\end{figure}

A few observations can be made regarding the fit values of $\alpha_{n}(T), C_{n}(T)$.  For both $\ell=1,2$ cases, $C_{n}$ is an increasing function of the temperature $T$ for both metric models.  The dependence of $\alpha_{n}$ on the $T$ is a bit more complicated.  For $\ell=2$ in both models, $\alpha_{n}(T)$ is a decreasing function of the temperature.  On the other hand for $\ell=1$, $\alpha_{n}(T)$ increases with $T$ until around $T=0.10$ GeV, then drops steadily as $T$ gets larger.  Since the deconfinement temperature $T_{c}$ is larger than $0.10$ GeV, we can say that in QGP phase, $\alpha_{n}(T)$, for each $\ell, n$, is a decreasing function of $T$.  This is in contrast to $\ell=0$ case where $\alpha_{n}$ is constant~(containing 't~Hooft coupling).  

In Fig.~\ref{12-fig}, the nature of increasing function of temperature $C_{n}(T)$ for $\ell=1,2$ appears consistent with the fact that for $\ell=0$, $C_{3,4}(T)\sim T^{2},~T$ respectively.  There is small deviation at large $T$ from the $\ell=0$ form due to rotation effects.  

\section{Conclusions and Discussions }

We have considered the effective potential between heavy quark and antiquark being submerged in the QGP for the states with $\ell=0,1,2$.  We have shown that even in conventional color-screened potential, introduction of the angular momentum barrier makes the excited states of quarkonium much less tightly bound, as well as making the potential at nonzero temperature crossing over zero value at finite distances.  We define the distance where the effective potential turns from negative to positive as distance grows as ``screening length'' $L^{*}$,  in contrast to the screening radius $r_{D}$ where the potential is suppressed exponentially but not exactly zero.  

The effective potentials for $\ell=0$ at zero and nonzero temperature at the leading order are derived analytically for general $n~(=3,4)$ in our setup, Eqn.~(\ref{eq:Vt0}),(\ref{eq:Vt})~(see also Appendix A).  The asymptotic expansion works well as long as $r<1/T$.  The mass dependence of the effective potential, Eqn.~(\ref{eq:Vumax}), is derived at the leading order together with its temperature dependence.  The leading order mass dependence is of $O(1/m^{2}),O(1/m^{3})$ for $n=3,4$ respectively.  This dependence remains even at zero temperature and depends on $r$ as $O(1/r^{6}),O(1/r^{4})$ for $n=3,4$.    

The numerical plots of the effective potential $V_{eff}(r,T)$ are given for each metric model with $\ell\equiv J=0,1,2$ at various temperature values.  An intriguing observation is that the position of the minimum $r_{0}$ of the effective potential is fixed once we fix the angular momentum, independent of the temperature.  Only angular momentum determines $r_{0}$ of the effective potential for each metric model.

Due to higher curvature, the $AdS_{5}$ metric model gives ``early-melting'' potential for states with high angular momentum.  As is shown in Ref.~\cite{bl}, while $\ell=2$ state of $n=3$ model resists to melting until $T=0.35$ GeV, $\ell=2$ of $n=4$ melts as early as $T=0.24$ GeV.  The shape of $V_{eff}$ for $\ell=2$ of $n=4$ metric model also shows very lightly bound potential at low temperature.  

The angular momentum of the quark antiquark system shows independence with respect to the temperature of the QGP, depending only on $\omega$ and $r$.  This can be understood that as long as the interaction between quark and antiquark calculated from the gravity dual models is radial, angular momentum of the bound state will not be altered as the temperature changes. 

Finally, fitting of the asymptotic form $V_{eff}=C_{n}(T)-\alpha_{n}(T)/r^{2/(n-2)}$~(motivated from $\ell=0$ case) to the potential of $\ell=1,2$ states in both metric models works very well for zero and nonzero temperature.  While the constant $C_{n}$ is an increasing function of temperature, the constant $\alpha_{n}$ is a decreasing function for $T>T_{c}$.                 

\section*{Acknowledgments}
\indent
O.A. and P.B. were supported in part by the U.S. Department of Energy under contract number DE-FG02-01ER41155.  J.L. was supported in part by the NSF Grant No. PHY-0340729.

\appendix

\section{asymptotic form of effective potential for $\ell=0$ }

Since $n$ could be related to $p$ of the D$p$-branes which act as the source generating the curved background metric of the gravity dual as $n=7-p$, it is useful to express the asymptotic form of the effective potential $V_{eff}(r,T)$ for $\ell=0$ in $u_{max}\to \infty$ limit, Eqn.~(\ref{eq:Vt}), as function of generic $n$,
\begin{eqnarray}
V_{eff}(r,T)& = & \frac{u_{h}}{\pi} - \frac{1}{\pi}\left(\frac{1}{2}-\frac{1}{n}\right)\frac{B_{n}a_{n}}{r^{2/(n-2)}}\left[1+O((rT)^{2n/(n-2)}) \right],
\end{eqnarray}
where 
\begin{eqnarray}
B_{n}& = & \left(\frac{2a_{n}R^{n/2}}{n} \right)^{2/(n-2)},
\end{eqnarray}
with 
\begin{eqnarray}
a_{n}=\frac{\Gamma(1-\frac{1}{n})\Gamma(\frac{1}{2})}{\Gamma(\frac{3}{2}-\frac{1}{n})}.
\end{eqnarray}

\newpage

\begin{figure}[tbp]
\centering
\includegraphics[height=3.5in, angle=0]{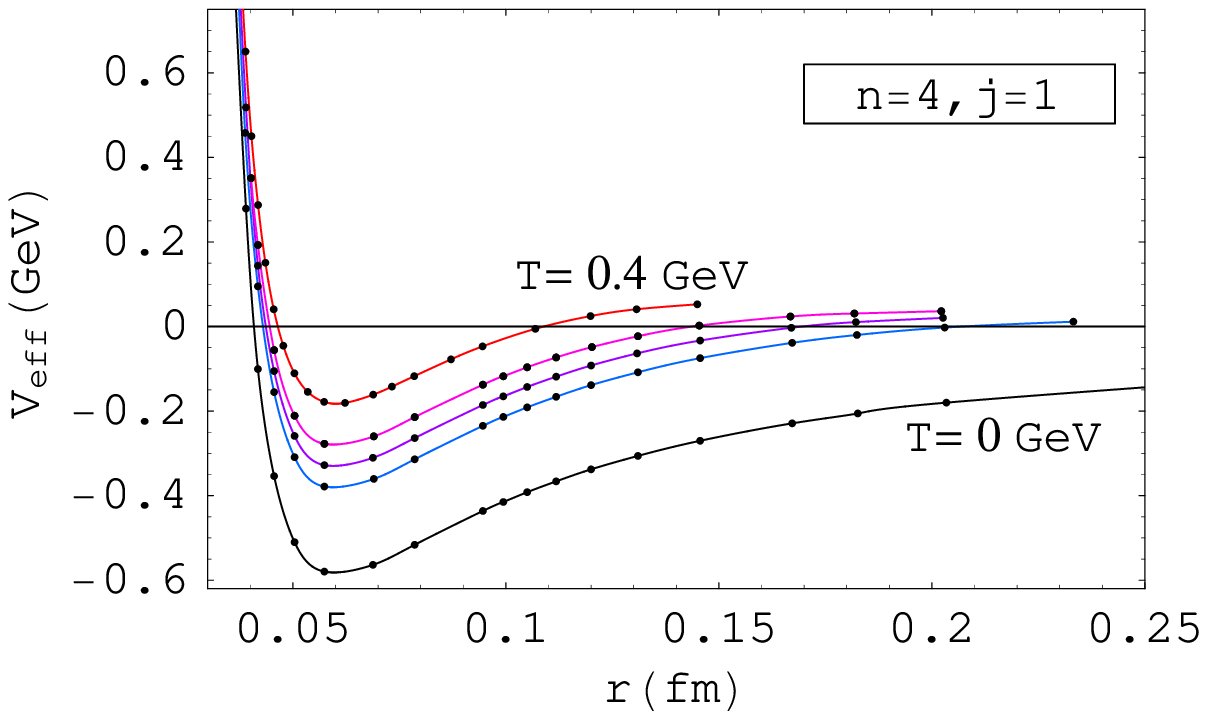}
\caption{$n=4, \ell \equiv J=1, T= 0,~0.20,~0.25,~0.30,~0.40$ GeV from bottom to top respectively. }
\label{3-fig}
\end{figure}

\begin{figure}[tbp]
\centering
\includegraphics[height=3.5in, angle=0]{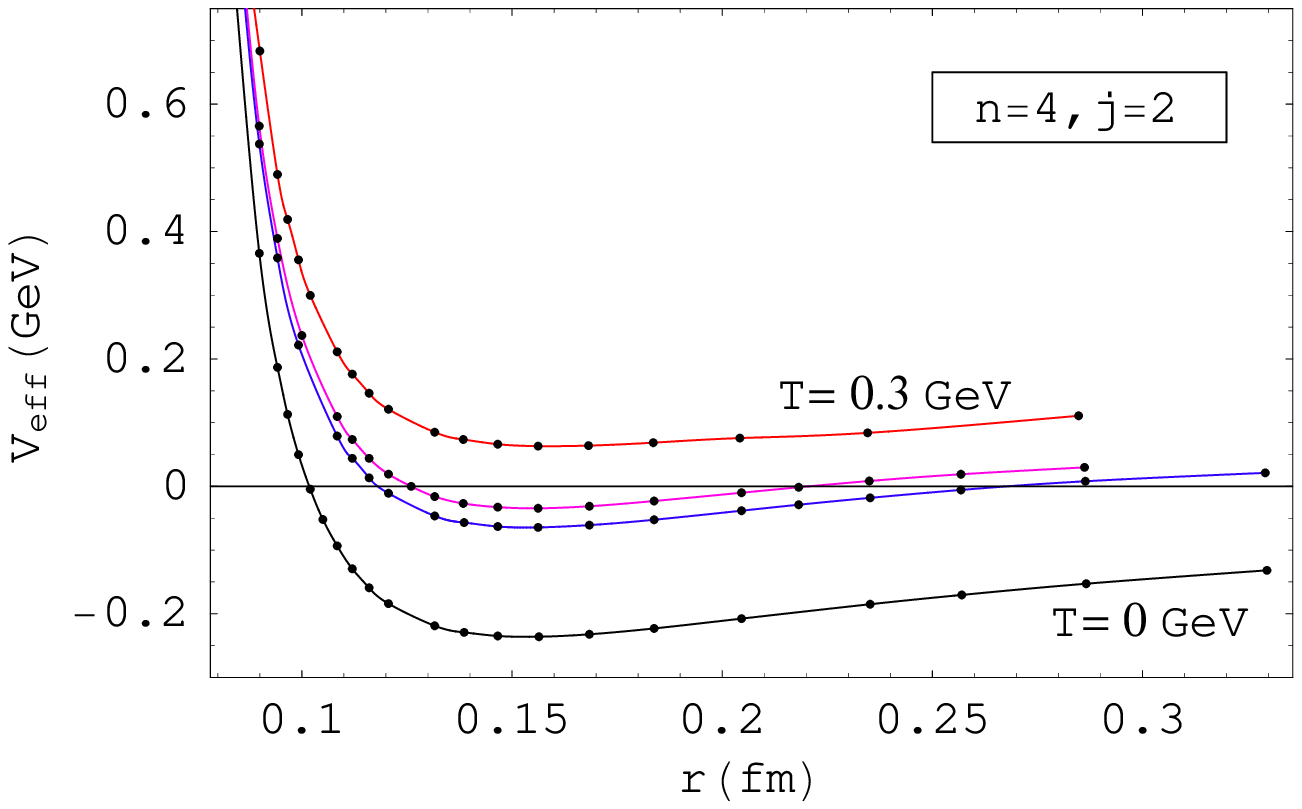}
\caption{$n=4, \ell \equiv J=2, T= 0,~0.17,~0.20,~0.30$ GeV from bottom to top respectively. }
\label{4-fig}
\end{figure}

\begin{figure}[tbp]
\centering
\includegraphics[height=3.5in, angle=0]{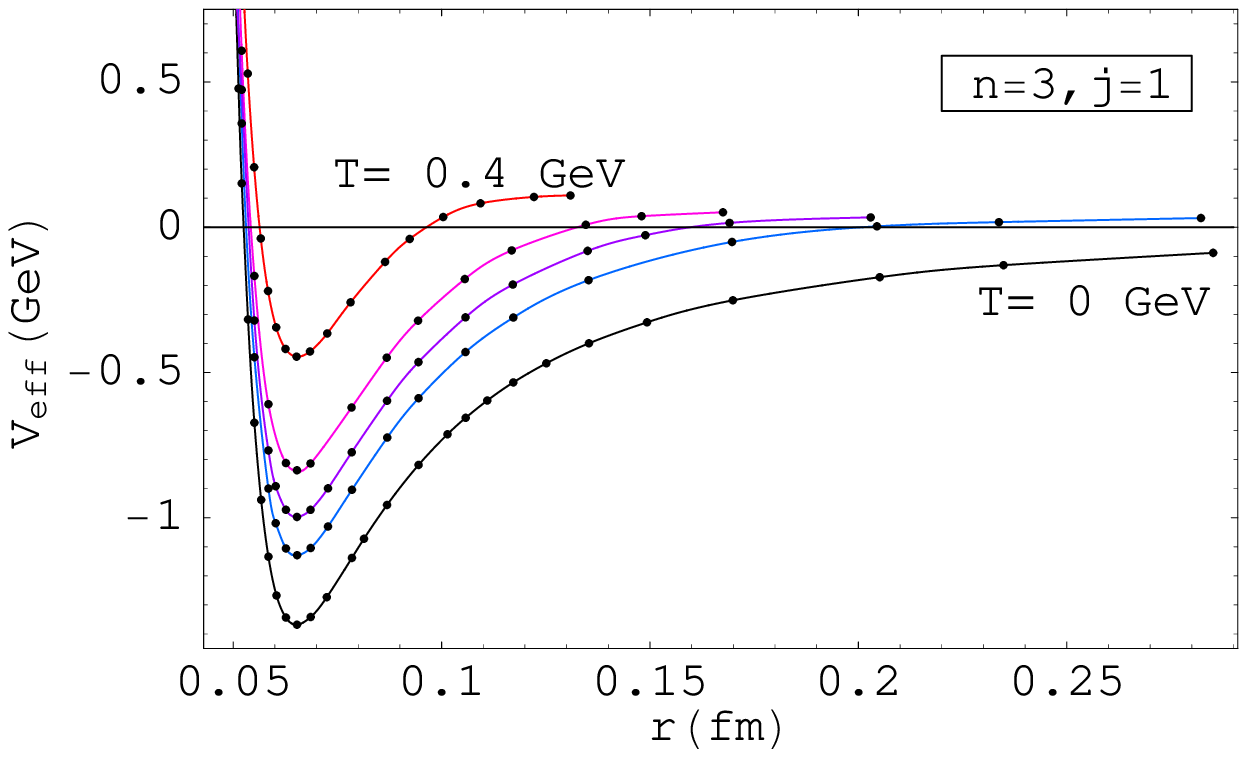}
\caption{$n=3, \ell \equiv J=1, T= 0,~0.20,~0.25,~0.30,~0.40$ GeV from bottom to top respectively. }
\label{5-fig}
\end{figure}

\begin{figure}[tbp]
\centering
\includegraphics[height=3.5in, angle=0]{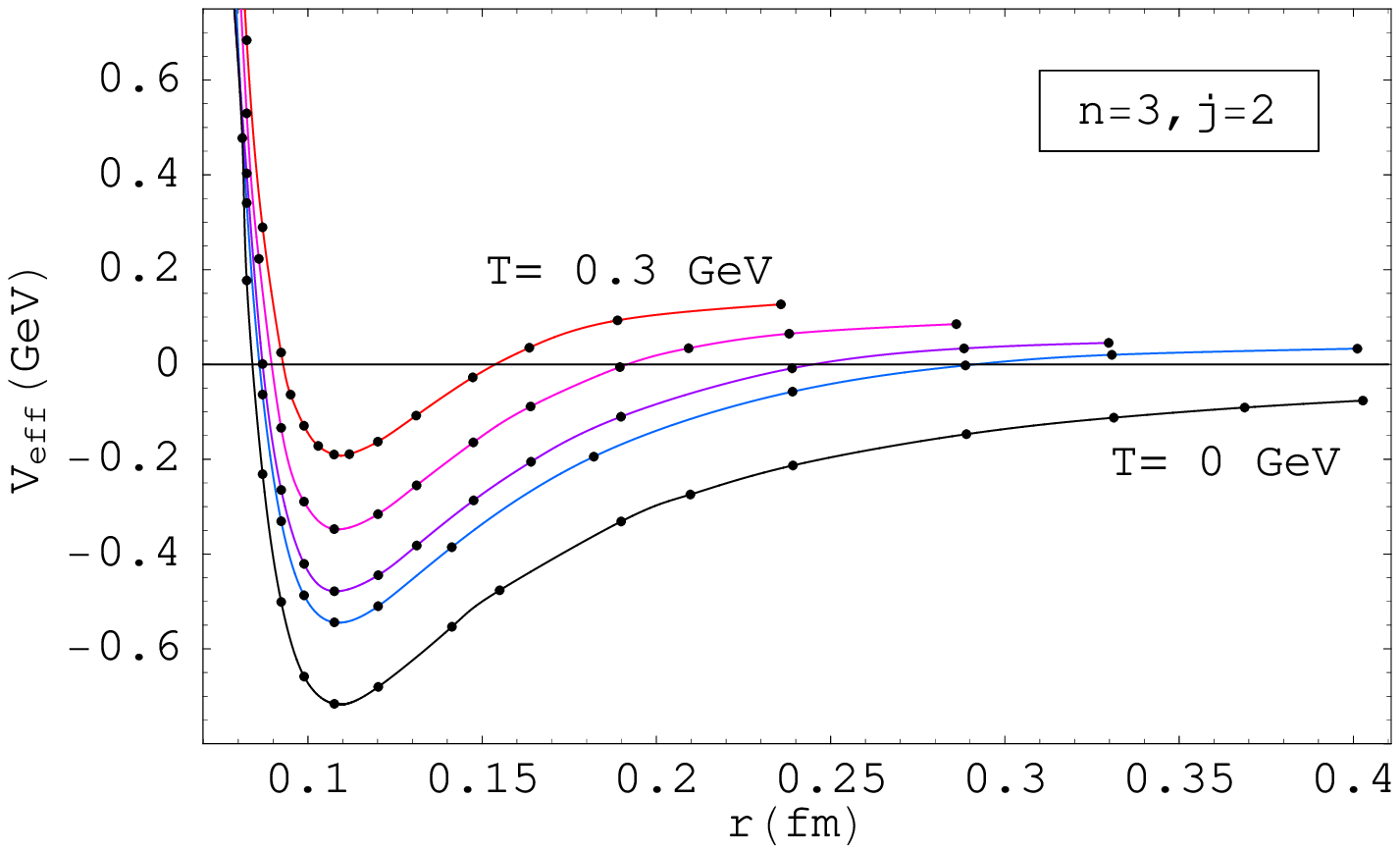}
\caption{$n=3, \ell \equiv J=2, T= 0,~0.17,~0.20,~0.25,~0.30$ GeV from bottom to top respectively. }
\label{6-fig}
\end{figure}

\end{document}